\documentclass[5p]{elsarticle}

\usepackage{amsmath,amssymb}
\usepackage{makecell}
\usepackage{graphicx}
\usepackage{dcolumn}
\usepackage{bm}

\usepackage[utf8]{inputenc}
\usepackage[T1]{fontenc}
\usepackage{mathptmx}
\usepackage{multirow}
\usepackage{float}
\usepackage{microtype}
\usepackage{stfloats}

\usepackage{array}
\newcolumntype{C}[1]{>{\centering\arraybackslash}p{#1}}

\begin{document}

\title{The Medium Amplitude Response of Nonlinear Maxwell-Oldroyd Type Models \\ in Simple Shear}

\author[1]{Kyle R. Lennon}
\author[2]{Gareth H. McKinley}
\author[1]{James W. Swan\corref{cor1}}%
\ead{jswan@mit.edu}
\cortext[cor1]{Corresponding author}
\address[1]{Department of Chemical Engineering, Massachusetts Institute of Technology, Cambridge, MA 02142}
\address[2]{Hatsopoulos Microfluids Laboratory, Department of Mechanical Engineering, Massachusetts Institute of Technology, Cambridge, MA 02139}

\date{\today}

\begin{abstract}

A general framework for Maxwell-Oldroyd type differential constitutive models is examined, in which an unspecified nonlinear function of the stress tensor and rate-of-deformation tensor is incorporated into the well-known corotational version of the Jeffreys model discussed by Oldroyd. For medium amplitude simple shear deformations, the recently developed mathematical framework of medium amplitude parallel superposition (MAPS) rheology reveals that this generalized nonlinear Maxwell model can produce only a limited number of distinct signatures, which combine linearly in a well-posed basis expansion for the third order complex viscosity. This basis expansion represents a library of MAPS signatures for distinct constitutive models that are contained within the generalized nonlinear Maxwell model. We describe a framework for quantitative model identification using this basis expansion for the third order complex viscosity, and discuss its limitations in distinguishing distinct nonlinear features of the underlying constitutive models from medium amplitude shear stress data. The leading order contributions to the normal stress differences are also considered, revealing that only the second normal stress difference provides distinct information about the weakly nonlinear response space of the model. After briefly considering the conditions for time-strain separability within the generalized nonlinear Maxwell model, we apply the basis expansion of the third order complex viscosity to derive the medium amplitude signatures of the model in specific shear deformation protocols. Finally, we use these signatures for estimation of model parameters from rheological data obtained by these different deformation protocols, revealing that three-tone oscillatory shear deformations produce data that is readily able to distinguish all features of the medium amplitude, simple shear response space of this generalized class of constitutive models.

\begin{keyword}
medium amplitude, Volterra series, Maxwell-Oldroyd type constitutive models, normal stresses
\end{keyword}

\end{abstract}

\maketitle

\section{Introduction}

Accurate constitutive modeling of viscoelastic materials is essential for understanding rheological data and extrapolating the measured behavior of a material from one deformation history to another. Constitutive viscoelastic models span a diverse range of forms \cite{rivlin-1971,bird-1976,bird-1987}, from integral models such as the K-BKZ family \cite{bernstein-1963} to differential models such as the 8-constant framework introduced in the seminal work of J.G. Oldroyd \cite{oldroyd-1958}. The merits of these constitutive models are varied; some are simple and easy to apply, some are general and able to describe the behavior of quite different materials, and some are linked directly to physical theories. As said by Oldroyd himself \cite{oldroyd-1950}, ``there may be many ways of generalizing the results of a single experiment, and the method of generalization so as to give a universally valid form of the equations of state is by no means obvious.'' In other words, identifying the constitutive model that best describes a given material is a difficult task. It involves first determining the model or family of models that is most appropriate for a material given the available data, then using the data to determine the optimal values of the model parameters. The latter task can be performed in a mathematically rigorous manner through different regression or inference techniques \cite{freund-2015}. The former, however, tends to be imprecise and relies heavily on expertise in the field, thus is prone to bias from personal experience or preference. For the sake of generality and future automated data-driven approaches, it is desirable to minimize the role that subjective model selection plays in model identification.

One way to minimize the role of subjectivity in model identification is to develop general frameworks which encompass more specific constitutive models. Subjective decisions are then reduced to determining the appropriate framework, rather than the individual model, and model selection can be performed mathematically therein. In linear viscoelasticity, for instance, the generalized Maxwell model encompasses a diverse set of linear viscoelastic behavior by coupling a linear arrangement of springs and dashpots with either a discrete set of relaxation modes, or a continuous distribution of relaxation times \cite{tschoegl-1989,lodge-1964}. Within this framework, material properties such as the complex modulus can be described by a (potentially infinite) basis expansion, called a Prony series, with each basis function corresponding to a discrete Maxwellian relaxation mode \cite{soussou-1970}. Though it may not be possible to associate each mode in the Prony series for a particular material with a distinct aspect of the material's microstructural physics, this expansion provides a convenient mathematical tool for accurately characterizing the linear viscoelastic behavior of many materials, often using only a relatively small number of adjustable parameters \cite{park-1999,park-2001}.

Various extensions to the generalized Maxwell model in nonlinear viscoelasticity exist. For example, the Lodge-type model proposed by Yamamoto allows the distribution of relaxation times to vary with the instantaneous strength of the deformation \cite{yamamoto-1971}. Fitting this shear rate-dependent relaxation time distribution to discrete rheological data, however, remains a difficult task, thus this model has not seen widespread use outside of theoretical studies. Other models of the Maxwell type are much simpler to apply, including many differential tensorial constitutive models. Because these models employ one of Oldroyd's objective derivatives to preserve frame-invariance, they are sometimes referred to as ``Maxwell-Oldroyd type'' models \cite{owens-2002}. One well-known model of this form is the Oldroyd 8-constant framework \cite{oldroyd-1958}, which encompasses common models such as the corotational Maxwell, Oldroyd B, and Johnson-Segalman models \cite{dewitt-1955,oldroyd-1950,johnson-1977,johnson-1981}. Within this framework, model selection equates to the purely quantitative task of parameter estimation. By its construction, however, the Oldroyd 8-constant framework excludes any differential constitutive models that possess terms that are nonlinear in the stress tensor. Notable examples of such models include the Giesekus and Phan-Thien--Tanner models, both of which have found success in modeling the rheology of many polymeric fluids \cite{giesekus-1982,tanner-1977}. Thus, even in the space of Maxwell-Oldroyd type models, the Oldroyd 8-constant framework is clearly not the most general framework possible.

Here, we study a more general framework for Maxwell-Oldroyd type constitutive models, which encapsulates all of the aforementioned differential constitutive models and many more. This framework, similar to that proposed previously by Leonov \cite{leonov-1992}, simply appends a general nonlinear function of the stress tensor and rate-of-deformation tensor to the frame-invariant corotational Jeffreys model \cite{bird-1987}. The details that distinguish individual nonlinear, Maxwell-Oldroyd type constitutive models lie entirely within this nonlinear function. 

Though this abstraction may seem to render analytical studies of this framework impossible, we are able to systematically study this general framework asymptotically. Examining the asymptotic nonlinearities of complex fluids and soft solids has recently gained popularity, in particular through a framework called medium amplitude oscillatory shear (MAOS) that studies the third-order asymptotically nonlinear response, commonly called the medium amplitude response, of materials undergoing single-tone oscillatory deformation in simple shear \cite{ewoldt-2013}. Recently, the framework for medium amplitude parallel superposition (MAPS) has been proposed to describe the medium amplitude response of materials under arbitrary simple shear deformations, including single- and multi-tone oscillatory protocols \cite{lennon-2020-1,lennon-2020-2}. As we show in the present paper, pairing the general framework for Maxwell-Oldroyd type constitutive models with MAPS rheology reveals all of the features that can possibly be observed in medium amplitude simple shear rheological studies of this class of models.

In this contribution, we will study this general framework for Maxwell-Oldroyd type constitutive models systematically using the tools of MAPS rheology. In Section \ref{sec:generalized_nonlinear_maxwell}, the framework will be precisely defined in terms of what we call the \emph{generalized nonlinear Maxwell model}. Section \ref{sec:maps_response} will then explore the medium amplitude signatures of this model in simple shear, both in the shear stress and normal stress responses. A critical step in this analysis is realizing that, in medium amplitude flows, we may construct a polynomial expansion for the unspecified nonlinear function in the generalized nonlinear Maxwell model, whose terms must obey the frame-invariant and isotropic properties expected from Oldroyd's principles of rheological invariance \cite{oldroyd-1950}. In Section \ref{sec:time_strain_separability}, we comment briefly on time-strain-separability in the context of this general framework.

The analysis in the first part of this work will reveal that the fusion of MAPS rheology and the generalized nonlinear Maxwell model results in a compact framework for model identification. In particular the stress response of the model to medium amplitude simple shear deformations can be represented as a basis expansion, and the coefficients of this expansion relate directly to material coefficients in the underlying constitutive model. Quantitative model identification is possible simply by determining these coefficients from data. It remains, however, to determine what data is best suited for this model identification problem. Sections \ref{sec:applied} and \ref{sec:model_identification} examine this question by applying the MAPS signatures determined in the previous sections to obtain asymptotic solutions for specific deformation protocols. We then perform simple numerical experiments using these analytical solutions to examine which experimental protocols produce data sets best suited for parameter estimation, in the sense that the uncertainty in the regressed parameters is minimal. This analysis simultaneously demonstrates the power of MAPS rheology as a framework for describing the material response to disparate flows, and for designing and interpreting experiments that produce data sets that are uniquely well-suited for the problem of model identification.

\section{The Generalized Nonlinear Maxwell Model}
\label{sec:generalized_nonlinear_maxwell}

In linear viscoelasticity, the Maxwell model is a well known and widely applied phenomenological constitutive model. The theoretical, single-mode Maxwell fluid contains two micromechanical elements, a Hookean spring and a Newtonian dashpot, which respond in series to imposed deformations \cite{tschoegl-1989}. The generalized Maxwell model is composed of a number of these single-mode Maxwell elements, which respond to imposed deformations in parallel. Each spring and each dashpot in this model may possess a distinct modulus of elasticity $E_k$ and viscosity $\eta_k$, respectively, thus the model may be represented by the set of equations:
\begin{subequations}
\begin{equation}
    \sigma_k + \tau_k \frac{d\sigma_k}{dt} = \eta_k \dot\gamma,
    \label{eq:linear_maxwell}
\end{equation}
\begin{equation}
    \sigma = \sum_k \sigma_k,
\end{equation}
\end{subequations}
where $\sigma_k$ represents the shear stress in the $k$th Maxwell mode, $\sigma$ is the total shear stress, $\dot\gamma$ is the shear strain rate, and $\tau_k = \eta_k/E_k$ is called the `relaxation time' of the $k$th mode. The complex viscosity for this linear model is compactly represented as a series of rational functions, each corresponding to the complex viscosity of a single Maxwell mode:
\begin{equation}
    \eta^*(\omega) = \sum_k \frac{\eta_k}{1 + i\tau_k\omega}.
    \label{eq:eta1_generalized}
\end{equation}
For materials that are well-represented by a small number of Maxwell modes, equation \ref{eq:eta1_generalized} is a convenient basis expansion for the complex viscosity as a function of the frequency $\omega$.

It is sometimes the case (e.g. especially for dilute solutions of a monodisperse polymer) that the linear viscoelastic response of a fluid is well-described by a single Maxwell element plus a Newtonian solvent (i.e. $\tau_2 = 0$). In this case, another compact representation of the total shear stress exists, called the Jeffreys model:
\begin{equation}
    \sigma + \lambda_1 \frac{d\sigma}{dt} = \eta_0\left(\dot\gamma + \lambda_2\frac{d\dot\gamma}{dt}\right),
    \label{eq:jeffreys_linear}
\end{equation}
where $\eta_0 = \eta_1 + \eta_2$, $\lambda_1 = \tau_1$ is still called the `relaxation time' of the single Jeffreys element, and $\lambda_2 = \lambda_1 \eta_2/\eta_0$ is called the `retardation time' of the Jeffreys element. In principle, multiple Jeffreys elements with distinct $\eta_{0,k}$, $\lambda_{1,k}$, and $\lambda_{2,k}$ can be combined in parallel to produce a generalized multi-mode model equivalent to the generalized Maxwell model with one mode having $\tau_i = 0$.

Adapting the linear Maxwell model to nonlinear, three-dimensional flows involves only a few modifications. First, the scalar shear stress elements $\sigma_k$ are replaced by tensorial contributions $\boldsymbol{\sigma}_k$ to the extra stress tensor $\boldsymbol{\sigma}$, which is equal to the total stress tensor $\boldsymbol{\pi}$ less an isotropic, hydrostatic pressure $p$:
\begin{equation}
    \boldsymbol{\pi} \equiv \boldsymbol{\sigma} - p\boldsymbol{\delta}, \quad \boldsymbol{\sigma} = \sum_k \boldsymbol{\sigma}_k,
\end{equation}
where $\boldsymbol{\delta}$ represents the identity tensor. Similarly, the scalar shear rate is replaced by the rate-of-deformation tensor $\boldsymbol{\dot\gamma}$:
\begin{equation}
    \boldsymbol{\dot\gamma} \equiv (\nabla \textbf{u}) + (\nabla \textbf{u})^{T},
\end{equation}
where $\textbf{u}$ is the velocity profile. Second, the scalar time derivative must be replaced by a materially-objective time derivative, which preserves invariance of the model to changes in the frame of reference. Objective time derivatives include the upper-convected derivative, the lower-convected derivative, and the corotational derivative \cite{bird-1987}. Here, we employ the corotational derivative:
\begin{equation}
    \frac{\mathcal{D}\textbf{x}}{\mathcal{D}t} \equiv \frac{D\textbf{x}}{Dt} + \frac{1}{2}(\boldsymbol{\omega}\cdot\textbf{x} - \textbf{x}\cdot\boldsymbol{\omega}),
\end{equation}
with $\boldsymbol{\omega}$ representing the vorticity tensor:
\begin{equation}
    \boldsymbol{\omega} \equiv (\nabla \textbf{u}) - (\nabla \textbf{u})^{T}.
\end{equation}
As a result of these changes to the linear, scalar Maxwell model (equation \ref{eq:linear_maxwell}), we arrive at the well-known corotational Maxwell model \cite{dewitt-1955}:
\begin{equation}
    \boldsymbol{\sigma}_k + \tau_k \frac{\mathcal{D}\boldsymbol{\sigma}_k}{\mathcal{D}t} = \eta_k\boldsymbol{\dot\gamma}.
\end{equation}
Similarly, by applying these generalizations to equation \ref{eq:jeffreys_linear}, we arrive at the related corotational Jeffreys model \cite{bird-1987}:
\begin{equation}
    \boldsymbol{\sigma}_k + \lambda_{1,k} \frac{\mathcal{D}\boldsymbol{\sigma}_k}{\mathcal{D}t} = \eta_{0,k}\left(\boldsymbol{\dot\gamma} + \lambda_{2,k}\frac{\mathcal{D}\boldsymbol{\dot\gamma}}{\mathcal{D}t}\right).
\end{equation}

Though the corotational Maxwell and Jeffreys models represent direct nonlinear extensions of the flexible linear generalized Maxwell framework, which is capable of modeling the linear viscoelastic behavior of a broad range of materials, in the space of nonlinear models they are quite limited. In particular, both the number of modes and all of the model parameters may be determined from the linear viscoelastic response, leaving no adjustable parameters to model distinct nonlinear phenomena. Thus, even for materials that display a linear viscoelastic response consistent with a single-mode Maxwell model, the corotational Maxwell and Jeffreys models have been shown to fail at accurately describing nonlinear behavior, for example by over-predicting the extent of shear thinning in steady flow curves \cite{gurnon-2012,lennon-2020-2}. 

Other forms have therefore been proposed to extend the linear Maxwell model to the nonlinear regime. Here, we focus solely on models that are represented by a set of differential equations, each of which is tensorial in nature and reduces to the linear Maxwell model for small deformations. The Oldroyd 8-constant framework is one such generalized model, which can reduce to either the corotational Maxwell or Jeffreys model with the appropriate parameter substitutions \cite{bird-1987}. However, the Oldroyd 8-constant framework is unable to capture many models of the Maxwell-Oldroyd form, such as the Giesekus and Phan-Thien--Tanner models. Despite its flexibility, therefore, it is clearly not the most general framework possible in this space.

A simple model that dramatically expands the flexibility of a Maxwell-Oldroyd type framework is one that adds to each mode of the corotational Jeffreys fluid a nonlinear function of the $k$th contribution to the extra stress tensor and the rate-of-deformation tensor:
\begin{equation}
    \boldsymbol{\sigma}_k + \lambda_{1,k} \frac{\mathcal{D}\boldsymbol{\sigma}_k}{\mathcal{D}t} + F_k(\boldsymbol{\sigma}_k,\boldsymbol{\dot\gamma}) = \eta_{0,k}\left(\boldsymbol{\dot\gamma} + \lambda_{2,k}\frac{\mathcal{D}\boldsymbol{\dot\gamma}}{\mathcal{D}t}\right).
    \label{eq:generalized_nonlinear_maxwell}
\end{equation}
We refer to this model as the \emph{generalized nonlinear Maxwell model}. The nonlinear function $F_k(\boldsymbol{\sigma}_k,\boldsymbol{\dot\gamma})$ can take on relatively simple forms, capturing models such as the Giesekus or Johnson-Segalman models, or more complex forms, such as that for the Rolie-Poly model \cite{graham-2003,likhtman-2003}. With modern data-analytics tools such as neural differential equations, it may even be possible to learn the form of $F_k(\boldsymbol{\sigma}_k,\boldsymbol{\dot\gamma})$ directly from data, without specifying any underlying structure for the function \cite{chen-2018,rackauckas-2020}.

The ability to distinguish different forms of $F_k(\boldsymbol{\sigma}_k,\boldsymbol{\dot\gamma})$ is contingent on the type of data available. For example, not all forms may be distinguishable with shear stress data alone in simple shear experiments. Moreover, the specific deformation protocol used to obtain rheological data may limit our ability to distinguish certain features of this nonlinear function. In the medium amplitude regime, where only the leading order nonlinear response of a material is considered, it is possible to resolve the issue of deformation protocol-specific data via studies using MAPS rheology \cite{lennon-2020-1}. Therefore, in this work we closely examine the MAPS response of the generalized nonlinear Maxwell model (eq. \ref{eq:generalized_nonlinear_maxwell}), beginning in the following section.

Although in this section the development of the generalized nonlinear Maxwell model has allowed for multiple response modes (each with time scale $\tau_k$), we will only consider the single-mode version of this model for the remainder of this work. Therefore, the subscript $k$ will be dropped in subsequent sections. Because the stress component generated by each mode is additive in the full extra stress tensor, the strain-controlled MAPS signatures also combine additively for the multi-mode case. Thus, it is straightforward to extend the following results to the multi-mode case.

\section{MAPS Response of the Generalized Nonlinear Maxwell Model}
\label{sec:maps_response}

\subsection{MAPS Rheology}

In MAPS rheology, the nonlinear functional relationship between the shear stress and shear strain rate in simple shear deformations is expanded as a truncated Volterra series in the frequency domain \cite{lennon-2020-1}:
\begin{align}
    \hat{\sigma}(\omega) = & \eta^*_1(\omega)\hat{\dot{\gamma}}(\omega) + \nonumber \\
    & \frac{1}{(2\pi)^2}\int\int\int_{-\infty}^{\infty}\eta^*_3(\omega_1,\omega_2,\omega_3)\hat{\dot{\gamma}}(\omega_1)\hat{\dot{\gamma}}(\omega_2)\hat{\dot{\gamma}}(\omega_3) \times \nonumber \\
    & \delta(\omega - \sum_{m=1}^3\omega_m)d\omega_1d\omega_2d\omega_3 + O(\hat{\dot{\gamma}}^5).
    \label{eq:MAPS}
\end{align}
The leading order response function, $\eta^*_1(\omega) \equiv \eta^*(\omega)$, is the familiar complex viscosity from linear viscoelasticity (see equation \ref{eq:eta1_generalized}). The response function that appears at third order with respect to the shear strain rate, $\eta^*_3(\omega_1,\omega_2,\omega_3)$, is called the third order complex viscosity. In equation \ref{eq:MAPS}, the shear stress has been written in terms of its Fourier transform:
\begin{equation}
    \hat{\sigma}(\omega) = \int_{-\infty}^{\infty}e^{-i\omega t}\sigma(t)dt,
\end{equation}
and likewise for the shear strain rate. An analogous expression exists in which the shear stress is written as an expansion in terms of the shear strain, in which case the linear and third order response functions are $G^*_1(\omega)$ and $G^*_3(\omega_1,\omega_2,\omega_3)$, which are called respectively the (linear) complex modulus and the third order complex modulus. Furthermore, the expansion can also be written for stress controlled experiments, in which the shear strain is expanded in terms of the shear stress. The appropriate response functions, $J^*_1(\omega)$ and $J^*_3(\omega_1,\omega_2,\omega_3)$, called the (linear) complex compliance and third order complex compliance, are related to the linear and third order complex moduli by a set of simple expressions. For more details on these representations of MAPS rheology, as well as important properties of the third order response functions, we refer readers to Lennon et. al \cite{lennon-2020-1}.

Analytical expressions for the third order complex viscosity and other MAPS functions often require sums and products over their three arguments: $(\omega_1,\omega_2,\omega_3)$. Throughout this work, we will employ the following shorthand notations for sums:
\begin{equation}
    \sum_j \Longleftrightarrow \sum_{j=1}^{3}, \qquad \sum_{k \neq j} \Longleftrightarrow \sum_{\substack{k=1\\k\ne j}}^3,
\end{equation}
and likewise for products.

\subsection{Cubic Expansion of the Generalized Nonlinear Maxwell Model}

In any medium amplitude study, only effects up to third order in the amplitude $\gamma_0$ of the imposed deformation are observed and/or described. Because both the rate-of-deformation and extra stress tensors are at least $O(\gamma_0)$ quantities, only the linear, quadratic, and cubic terms in a constitutive model are apparent in the medium amplitude regime. Thus, only the cubic expansion of $F(\boldsymbol{\sigma},\boldsymbol{\dot\gamma})$ is identifiable in medium amplitude experiments. If we also specify the constraint that the model is isotropic, then there are only a finite number of terms in such an expansion; thus, it is possible to enumerate all features of $F(\boldsymbol{\sigma},\boldsymbol{\dot\gamma})$ that are apparent in MAPS rheology.

Because $F(\boldsymbol{\sigma},\boldsymbol{\dot\gamma})$ is a tensor-valued function of two tensors, and is furthermore expected to obey frame-invariance in order for the model to be physically meaningful, its polynomial expansion must be written with care. Spencer and Rivlin \cite{spencer-1958} developed a useful theory for matrix polynomials of one or more tensorial quantities, which we apply here. In essence, the theory states that each term in the polynomial should be some scalar coefficient multiplying one or more inner products of the tensorial arguments (in this case $\boldsymbol{\sigma}$ and $\boldsymbol{\dot\gamma}$) and the identity tensor $\boldsymbol{\delta}$. The coefficients themselves should subsequently be written as polynomial expansions in terms of the invariants of each term in the polynomial expansion, for example $I(\boldsymbol{\sigma}) = \mathrm{tr}(\boldsymbol{\sigma}) = \sum_{i=1}^3 \sigma_{ii}$ or $II(\boldsymbol{\sigma}) = (\mathrm{tr}(\boldsymbol{\sigma})^2 - \mathrm{tr}(\boldsymbol{\sigma}^2))/2$. Because the higher invariants of any term are related to the first invariants of that term and higher-order terms (e.g. $II(\boldsymbol{\sigma}) = (I(\boldsymbol{\sigma})^2 - I(\boldsymbol{\sigma}\cdot\boldsymbol{\sigma}))/2$) by the Cayley-Hamilton theorem, it is sufficient to only consider the first invariant, i.e. the trace. Moreover, we focus on incompressible materials in simple shear, for which $\mathrm{tr}(\boldsymbol{\dot\gamma})$ is zero and the normal stress contributions that form $\mathrm{tr}(\boldsymbol{\sigma})$ are necessarily second order in the deformation amplitude. In this manner, it is possible to systematically enumerate each possible term in the expansion of $F(\boldsymbol{\sigma},\boldsymbol{\dot\gamma})$ up to cubic order in the deformation amplitude. The result is an isotropic, frame-invariant differential constitutive model that we call the \emph{cubic Maxwell model}, which is presented in equation \ref{eq:cubic_model}.

\begin{figure*}
\begin{align}
    & \boldsymbol{\sigma} + \lambda_{1}\frac{\mathcal{D}\boldsymbol{\sigma}}{\mathcal{D}t} + \frac{1}{2}\mu_{0}(\text{tr}\boldsymbol{\sigma})\boldsymbol{\dot{\gamma}} - \frac{1}{2}\mu_{1}\{\boldsymbol{\sigma}\cdot\boldsymbol{\dot{\gamma}} + \boldsymbol{\dot{\gamma}}\cdot\boldsymbol{\sigma}\} +
    \frac{1}{2}\nu_{1}(\boldsymbol{\sigma}:\boldsymbol{\dot{\gamma}})\boldsymbol{\delta} - \frac{\alpha_0}{\eta_0}(\text{tr}\boldsymbol{\sigma})\boldsymbol{\sigma} + \frac{\alpha_1}{\eta_0}\{\boldsymbol{\sigma}\cdot\boldsymbol{\sigma}\} + \frac{\beta_0}{\eta_0}(\mathrm{tr}\boldsymbol{\sigma})^2\boldsymbol{\delta} -  \frac{\beta_1}{2\eta_0}(\boldsymbol{\sigma}:\boldsymbol{\sigma})\boldsymbol{\delta} \nonumber \\
    & - \frac{c_1 \lambda_1}{\eta_0}(\boldsymbol{\sigma}:\boldsymbol{\dot\gamma})\boldsymbol{\sigma} - c_2\lambda_1(\boldsymbol{\dot\gamma}:\boldsymbol{\dot\gamma})\boldsymbol{\sigma} - \frac{c_3 \lambda_1}{\eta_0^2}(\boldsymbol{\sigma}:\boldsymbol{\sigma})\boldsymbol{\sigma} - \frac{d_1 \lambda_1}{\eta_0}(\boldsymbol{\sigma}:\boldsymbol{\sigma})\boldsymbol{\dot\gamma} - d_2\lambda_1(\boldsymbol{\sigma}:\boldsymbol{\dot\gamma})\boldsymbol{\dot\gamma} \nonumber \\
    & - \frac{f_1\lambda_1}{\eta_0}\{\boldsymbol{\sigma}\cdot\boldsymbol{\sigma}\cdot\boldsymbol{\dot\gamma} + \boldsymbol{\dot\gamma}\cdot\boldsymbol{\sigma}\cdot\boldsymbol{\sigma}\} - f_2\lambda_1\{\boldsymbol{\sigma}\cdot\boldsymbol{\dot\gamma}\cdot\boldsymbol{\dot\gamma} + \boldsymbol{\dot\gamma}\cdot\boldsymbol{\dot\gamma}\cdot\boldsymbol{\sigma}\} - \frac{2f_3\lambda_1}{\eta_0^2}\{\boldsymbol{\sigma}\cdot\boldsymbol{\sigma}\cdot\boldsymbol{\sigma}\} + \zeta (\mathrm{tr}\boldsymbol{\sigma})\boldsymbol{\delta} \nonumber \\
    & \qquad\qquad\qquad\qquad\qquad = \eta_{0}\left(\boldsymbol{\dot{\gamma}} + \lambda_{2}\frac{\mathcal{D}\boldsymbol{\dot{\gamma}}}{\mathcal{D}t} - \mu_{2}\{\boldsymbol{\dot{\gamma}}\cdot\boldsymbol{\dot{\gamma}}\} + \frac{1}{2}\nu_{2}(\boldsymbol{\dot{\gamma}}:\boldsymbol{\dot{\gamma}})\boldsymbol{\delta} + d_3\lambda_1(\boldsymbol{\dot\gamma}:\boldsymbol{\dot\gamma})\boldsymbol{\dot\gamma} + 2f_4\lambda_1\{\boldsymbol{\dot\gamma}\cdot\boldsymbol{\dot\gamma}\cdot\boldsymbol{\dot\gamma}\}\right).
    \label{eq:cubic_model}
\end{align}
\begin{align}
    \boldsymbol{\sigma} + \lambda_{1}\frac{\mathcal{D}\boldsymbol{\sigma}}{\mathcal{D}t} + \frac{1}{2}\mu_{0}(\text{tr}\boldsymbol{\sigma})\boldsymbol{\dot{\gamma}} - \frac{1}{2}\mu_{1}\{\boldsymbol{\sigma}\cdot\boldsymbol{\dot{\gamma}} + \boldsymbol{\dot{\gamma}}\cdot\boldsymbol{\sigma}\} +
    \frac{1}{2}\nu_{1}(\boldsymbol{\sigma}:\boldsymbol{\dot{\gamma}})\boldsymbol{\delta} & - \frac{\alpha_0}{\eta_0}(\text{tr}\boldsymbol{\sigma})\boldsymbol{\sigma} + \frac{\alpha_1}{\eta_0}\{\boldsymbol{\sigma}\cdot\boldsymbol{\sigma}\} + \frac{\beta_0}{\eta_0}(\mathrm{tr}\boldsymbol{\sigma})^2\boldsymbol{\delta} -  \frac{\beta_1}{2\eta_0}(\boldsymbol{\sigma}:\boldsymbol{\sigma})\boldsymbol{\delta} \nonumber \\
    & = \eta_{0}\left(\boldsymbol{\dot{\gamma}} + \lambda_{2}\frac{\mathcal{D}\boldsymbol{\dot{\gamma}}}{\mathcal{D}t} - \mu_{2}\{\boldsymbol{\dot{\gamma}}\cdot\boldsymbol{\dot{\gamma}}\} + \frac{1}{2}\nu_{2}(\boldsymbol{\dot{\gamma}}:\boldsymbol{\dot{\gamma}})\boldsymbol{\delta}\right).
    \label{eq:quadratic_model}
\end{align}
\makebox[\linewidth]{\rule{\textwidth}{0.4pt}}
\end{figure*}

The cubic expansion of $F(\boldsymbol{\sigma},\boldsymbol{\dot\gamma})$ contains twenty distinct terms that may appear at third order in simple shear. In equation \ref{eq:cubic_model}, the coefficients of these terms are constructed to be consistent with the nomenclature of the Oldroyd 8-constant framework, and such that every parameter has dimensions of time except for $\eta_0$, which has dimensions of (stress$\times$time), and $\zeta$, which is dimensionless. We have also employed the shorthand notation of the double dot product: $\textbf{A}:\textbf{B} \equiv \mathrm{tr}(\textbf{A}\cdot\textbf{B})$.

With the appropriate parameter assignments, equation \ref{eq:cubic_model} can be reduced to many different constitutive models. These include the Oldroyd 8-constant framework and all sub-models contained within it. In fact, Oldroyd in his 1958 paper originally proposed the 8-constant framework as a similar expansion of a general nonlinear function $F(\boldsymbol{\sigma},\boldsymbol{\dot\gamma})$, but included only terms that were linear in the rate-of-deformation tensor and quadratic overall \cite{oldroyd-1958}. Therefore, the leading order nonlinearities produced by models such as the Giesekus and Phan-Thien--Tanner models, which are not included in the Oldroyd 8-constant framework, are incorporated within the cubic Maxwell model. Other models, for which $F(\boldsymbol{\sigma},\boldsymbol{\dot\gamma})$ is not a polynomial, are approximated exactly to third order by the general cubic Maxwell model in simple shear deformations. These include both the stretching and non-stretching Rolie-Poly models, among others. An extensive list of models that are either approximated exactly to third order by, or completely contained within the cubic Maxwell model is presented in Table \ref{tab:sub_models}, along with the appropriate parameter assignments. For instance, equation \ref{eq:cubic_model} reduces to the Oldroyd 8-constant framework when all parameters except $\mu_0$, $\mu_1$, $\nu_1$, $\mu_2$, $\nu_2$, $\eta_0$, $\lambda_1$, and $\lambda_2$ are set to zero. Note that, for many of the models listed in Table \ref{tab:sub_models}, equation \ref{eq:cubic_model} with the appropriate parameter assignments reflects the non-Newtonian extra stress tensor, which is typically added to a Newtonian solvent contribution to produce the extra stress. Though it is possible to capture Newtonian solvent effects within equation \ref{eq:cubic_model} via the retardation term, we leave $\lambda_2 = 0$ for many such models in order to reflect their familiar forms.  Also, in Table \ref{tab:sub_models}, all models (aside from the Second Order Fluid) have nonzero $\eta_0$ and $\lambda_1$, but many of the other parameters in equation \ref{eq:cubic_model} are set to zero. Therefore, only the remaining non-zero parameters are listed.

\begin{table*}[ht!]
    \centering
    \begin{tabular}{|c|c|} \hline
        Model & Non-zero Parameters (excl. $\eta_0$, $\lambda_1$) \\[5pt] \hline\hline
        \multicolumn{2}{|c|}{Asymptotically Approximated by the Cubic Maxwell Model (eq. \ref{eq:cubic_model})} \\[5pt] \hline\hline
        Non-stretching Gaussian Rolie-Poly \cite{likhtman-2003} & $\mu_1 = \lambda_1$, $\nu_1 = \frac{2}{3}\lambda_1$, $c_1 = -\frac{1}{3}(1 + \beta_R)\lambda_1$ \\[5pt] \hline
        Stretching Gaussian Rolie-Poly \cite{likhtman-2003} & $\zeta = Z_R$, $\mu_1 = \lambda_1$, $\alpha_0 = -Z_R(1 + \beta_R)\lambda_1$, $\beta_0 = -\frac{1}{4}Z_R\lambda_1$ \\[5pt] \hline
        Stretching Gaussian cDCR-CS \cite{marrucci-2003} & $\zeta=\frac{1}{3}(3Z_c - 1)$, $\mu_1 = \lambda_1$, $\alpha_0 = -\frac{1}{3}\beta_c(3Z_c - 1)\lambda_1$, $\beta_0 = -\frac{1}{9}\beta_c(3Z_c - 1)\lambda_1$ \\[5pt] \hline\hline
        \multicolumn{2}{|c|}{Contained Within the Cubic Maxwell Model (eq. \ref{eq:cubic_model})} \\[5pt] \hline\hline
        Larson \cite{larson-1984} & $\mu_1 = \lambda_1$, $\nu_1 = \frac{2}{3}\xi\lambda_1$, $c_1 = -\frac{1}{3}\xi\lambda_1$ \\[5pt] \hline\hline
        \multicolumn{2}{|c|}{Contained Within the Quadratic Maxwell Model (eq. \ref{eq:quadratic_model})} \\[5pt] \hline\hline
        Oldroyd 8-Constant \cite{oldroyd-1958} & $\mu_0$, $\mu_1$, $\nu_1$, $\mu_2$, $\nu_2$, $\lambda_2$ \\[5pt] \hline
        Oldroyd 6-Constant \cite{oldroyd-1958} & $\mu_0$, $\mu_1$, $\mu_2$, $\lambda_2$ \\[5pt] \hline
        Oldroyd 4-Constant \cite{oldroyd-1958} & \begin{tabular}{c}
            $\mu_1 = \lambda_1$, $\mu_2 = \lambda_2$, $\mu_0$
        \end{tabular} \\[5pt] \hline
        Gordon-Schowalter \cite{gordon-1972} & \begin{tabular}{c}
            $\lambda_2 = \frac{\eta_s}{\eta_0}\lambda_1$, $\mu_1 = (1 - \xi)\lambda_1$, $\mu_2 = (1-\xi)\frac{\eta_s}{\eta_0}\lambda_1$
        \end{tabular} \\[5pt] \hline
        Johnson-Segalman \cite{johnson-1977,johnson-1981} & \begin{tabular}{c}
            $\mu_1 = (1 - \xi)\lambda_1$
        \end{tabular} \\[5pt] \hline
        Oldroyd Fluid A \cite{saengow-2017} & \begin{tabular}{c}
            $\mu_1 = -\lambda_1$, $\mu_2 = -\lambda_2$
        \end{tabular} \\[5pt] \hline
        Oldroyd Fluid B \cite{oldroyd-1950} & \begin{tabular}{c}
            $\mu_1 = \lambda_1$, $\mu_2 = \lambda_2$
        \end{tabular} \\[5pt] \hline
        Second Order Fluid \cite{bird-1987} & \begin{tabular}{c}
            $\mu_2$, $\lambda_2$, $\lambda_1 = 0$
        \end{tabular} \\[5pt] \hline
        Arb. Normal Stress Ratio (ANSR) \cite{saengow-2017} & \begin{tabular}{c}
            $\mu_1$
        \end{tabular} \\[5pt] \hline
        Corotational Jeffreys \cite{bird-1987} & \begin{tabular}{c}
            $\lambda_2$
        \end{tabular} \\[5pt] \hline
        Williams 3-Constant \cite{williams-1962} & \begin{tabular}{c}
            $\mu_1 = \lambda_1$, $\mu_2 = \lambda_2$, $\nu_1 = \frac{2}{3}\lambda_1$, $\nu_2 = \frac{2}{3}\lambda_2$
        \end{tabular} \\[5pt] \hline
        Denn Modified Maxwell \cite{denn-1971} & \begin{tabular}{c}
            $\mu_1 = \lambda_1$, $\mu_2$
        \end{tabular} \\[5pt] \hline
        Corotational Maxwell \cite{dewitt-1955} & \begin{tabular}{c}
            --
        \end{tabular} \\[5pt] \hline
        Upper Convected Maxwell \cite{bird-1987} & \begin{tabular}{c}
            $\mu_1 = \lambda_1$
        \end{tabular} \\[5pt] \hline
        Lower Convected Maxwell \cite{bird-1987} & \begin{tabular}{c}
            $\mu_1 = -\lambda_1$
        \end{tabular} \\[5pt] \hline
        Giesekus \cite{giesekus-1982} & \begin{tabular}{c}
            $\mu_1 = \lambda_1$, $\alpha_1 = \alpha\lambda_1$
        \end{tabular} \\[5pt] \hline
        Linearized Phan-Thien--Tanner \cite{tanner-1977} & \begin{tabular}{c}
            $\mu_1 = (1 - \xi)\lambda_1$, $\alpha_0 = -\epsilon \lambda_1$
        \end{tabular} \\[5pt] \hline
    \end{tabular}
    \caption{Models included in or approximated by the cubic Maxwell model (equation \ref{eq:cubic_model}). All parameters besides $\eta_0$ and $\lambda_1$ that are non-zero in a model are listed, along with any necessary parameter assignments. All other (unlisted) parameters in each row entry are identically zero.}
    \label{tab:sub_models}
\end{table*}

In many of the models listed in Table \ref{tab:sub_models}, the coefficients of all cubic terms -- those represented by Latin symbols in equation \ref{eq:cubic_model} -- are set to zero. Therefore, retaining only the quadratic terms still results in an apparently quite general model. Moreover, for all of those models which contain only quadratic terms, the coefficient $\zeta$ is zero as well. As we will find shortly, excluding this coefficient results in an especially simple form for the solution to the third order complex viscosity. Therefore, it is useful to define a more compact model, called the \emph{quadratic Maxwell model}, which is presented in equation \ref{eq:quadratic_model}.

The quadratic Maxwell model contains only eight adjustable parameters multiplying nonlinear terms, but is still able to capture a wide range of behaviors. As demonstrated in Table \ref{tab:sub_models}, the Oldroyd 8-constant framework, and all models represented therein, as well as the Giesekus and linearized Phan-Thien--Tanner models, are all exactly captured by the quadratic Maxwell model.

\subsection{MAPS Response of the Generalized Nonlinear Maxwell Model}
\label{sec:maps_solution}

Because only terms up to cubic overall order in the extra stress and rate-of-deformation tensor in a constitutive model influence the medium amplitude response of the model in simple shear, the MAPS response of the generalized nonlinear Maxwell model (equation \ref{eq:generalized_nonlinear_maxwell}) is the same as that of the cubic Maxwell model truncation (equation \ref{eq:cubic_model}). Thus, while it may not be obvious that equation \ref{eq:generalized_nonlinear_maxwell} possesses an analytical solution in any nonlinear flow, asymptotic solutions can indeed be derived using equation \ref{eq:cubic_model}.

Every term in the cubic Maxwell model contains either the extra stress tensor, the rate-of-deformation tensor, or both. In simple shear, the invariants of the rate-of-deformation tensor are set by the shear rate. Therefore, it is most convenient to obtain the solution for one of the two MAPS response functions that relate the shear stress to the shear rate -- either the third order complex fluidity $\phi^*_3(\omega_1,\omega_2,\omega_3)$ in stress control or the third order complex viscosity $\eta^*_3(\omega_1,\omega_2,\omega_3)$ in strain (rate) control. Previously, we noted that in the multi-mode generalization of the model, the strain-controlled MAPS response functions for each mode are additive. This is because, in strain control, the extra stress in the multi-mode cubic Maxwell model simply represents a sum of solutions to independent and linearly independent differential equations. In stress control, however, the multi-mode model represents a system of differential equations coupled by an algebraic equation. Therefore, for future ease in extending the results of the single-mode case to the multi-mode case, we will focus on the solution of the third order complex viscosity.

In strain-controlled MAPS rheology, we study the response of the model to a simple shear deformation with the shear rate $\dot\gamma(t) = \gamma_0 s(t)$, where $\gamma_0$ represents the characteristic amplitude of the deformation. The velocity profile is therefore:
\begin{equation}
    \textbf{u} = \gamma_0 s(t) x_2 \textbf{e}_1,
\end{equation}
where $\textbf{e}_j$ represents a unit vector in the $j$th direction, with $j = 1$ here representing the direction of the flow, and $x_2$ representing the position in the velocity gradient direction. To obtain the analytic form of the third order complex viscosity, we write the extra stress tensor as a power series in $\gamma_0$:
\begin{equation}
    \boldsymbol{\sigma} = \gamma_0 \boldsymbol{\sigma}^{(1)} + \gamma_0^2 \boldsymbol{\sigma}^{(2)} + \gamma_0^3 \boldsymbol{\sigma}^{(3)} + O(\gamma_0^4).
\end{equation}
This expression is substituted into equation \ref{eq:cubic_model}, resulting in a differential equation for each component of $\boldsymbol{\sigma}$. We then perform asymptotic analysis in the limit that $\gamma_0 \rightarrow 0$, grouping terms of like order in $\gamma_0$. At first order, we recover the linear Jeffreys model for the shear stress component, $\sigma^{(1)}_{12}$, from which we obtain the (linear) complex viscosity of the generalized nonlinear Maxwell model:
\begin{equation}
    \eta^*_1(\omega) = \eta_0\left(\frac{1 + i\lambda_2 \omega}{1 + i\lambda_1\omega}\right).
    \label{eq:eta1}
\end{equation}
\begin{figure*}[ht!]
\begin{equation}
    \frac{\eta^*_3(\omega_1,\omega_2,\omega_3)}{\eta_0} = \sum_{n=1}^{10}\left[a_{n}^{(0)}\Omega_n(\omega_1,\omega_2,\omega_3;0,0) + a_{n}^{(3)}\Omega_n(\omega_1,\omega_2,\omega_3;3,0) + b_{n}\Omega_n(\omega_1,\omega_2,\omega_3;3,1)\right].
    \label{eq:cubic_solution}
\end{equation}
\begin{subequations}
\begin{equation}
    \Omega_1(\omega_1,\omega_2,\omega_3;r,s) = \frac{1}{3}\left(\frac{1}{1 + i\lambda_1\underset{j}{\sum}\omega_j}\right)\left\{\sum_{j}\left[\left(\frac{1}{1 + r\zeta + i\lambda_1\underset{k\neq j}{\sum}\omega_k}\right)\left(\frac{1}{1 + i\lambda_1 s\underset{k\neq j}{\sum}\omega_k}\right)\right]\right\},
\end{equation}
\begin{equation}
    \Omega_2(\omega_1,\omega_2,\omega_3;r,s) = \frac{1}{6}\left(\frac{1}{1 + i\lambda_1\underset{j}{\sum}\omega_j}\right)\left\{\sum_{j}\left[\left(\frac{1}{1 + r\zeta + i\lambda_1\underset{k\neq j}{\sum}\omega_k}\right)\left(\frac{1}{1 + i\lambda_1 s\underset{k\neq j}{\sum}\omega_k}\right)\sum_{k\neq j}\left(\frac{1 + i\lambda_2\omega_k}{1 + i\lambda_1\omega_k}\right)\right]\right\},
\end{equation}
\begin{equation}
    \Omega_3(\omega_1,\omega_2,\omega_3;r,s) = \frac{1}{3}\left(\frac{1}{1 + i\lambda_1\underset{j}{\sum}\omega_j}\right)\left\{\sum_{j}\left[\left(\frac{1 + i\lambda_2\omega_j}{1 + i\lambda_1\omega_j}\right)\left(\frac{1}{1 + r\zeta + i\lambda_1\underset{k\neq j}{\sum}\omega_k}\right)\left(\frac{1}{1 + i\lambda_1 s\underset{k\neq j}{\sum}\omega_k}\right)\right]\right\},
\end{equation}
\begin{align}
    \Omega_4(\omega_1,\omega_2,\omega_3;r,s) = \frac{1}{6}&\left(\frac{1}{1 + i\lambda_1\underset{j}{\sum}\omega_j}\right)\times \\
    & \left\{\sum_{j}\left[\left(\frac{1 + i\lambda_2\omega_j}{1 + i\lambda_1\omega_j}\right)\left(\frac{1}{1 + r\zeta + i\lambda_1\underset{k\neq j}{\sum}\omega_k}\right)\left(\frac{1}{1 + i\lambda_1 s\underset{k\neq j}{\sum}\omega_k}\right)\sum_{k\neq j}\left(\frac{1 + i\lambda_2\omega_k}{1 + i\lambda_1\omega_k}\right)\right]\right\},
\end{align}
\begin{equation}
    \Omega_5(\omega_1,\omega_2,\omega_3;r,s) = \frac{1}{3}\left(\frac{1}{1 + i\lambda_1\underset{j}{\sum}\omega_j}\right)\left\{\sum_{j}\left[\left(\frac{1}{1 + r\zeta + i\lambda_1\underset{k\neq j}{\sum}\omega_k}\right)\left(\frac{1}{1 + i\lambda_1 s\underset{k\neq j}{\sum}\omega_k}\right)\prod_{k\neq j}\left(\frac{1 + i\lambda_2\omega_k}{1 + i\lambda_1\omega_k}\right)\right]\right\},
\end{equation}
\begin{equation}
    \Omega_6(\omega_1,\omega_2,\omega_3;r,s) = \frac{1}{3}\left(\frac{1}{1 + i\lambda_1\underset{j}{\sum}\omega_j}\right)\left[\prod_j\left(\frac{1 + i\lambda_2\omega_j}{1 + i\lambda_1\omega_j}\right)\right]\left\{\sum_{j}\left[\left(\frac{1}{1 + r\zeta + i\lambda_1\underset{k\neq j}{\sum}\omega_k}\right)\left(\frac{1}{1 + i\lambda_1 s\underset{k\neq j}{\sum}\omega_k}\right)\right]\right\},
\end{equation}
\begin{equation}
    \Omega_7(\omega_1,\omega_2,\omega_3;r,s) = \left(\frac{1}{1 + i\lambda_1\underset{j}{\sum}\omega_j}\right),
\end{equation}
\begin{equation}
    \Omega_8(\omega_1,\omega_2,\omega_3;r,s) = \frac{1}{3}\left(\frac{1}{1 + i\lambda_1\underset{j}{\sum}\omega_j}\right)\left[\sum_j\left(\frac{1 + i\lambda_2\omega_j}{1 + i\lambda_1\omega_j}\right)\right],
\end{equation}
\begin{equation}
    \Omega_9(\omega_1,\omega_2,\omega_3;r,s) = \frac{1}{3}\left(\frac{1}{1 + i\lambda_1\underset{j}{\sum}\omega_j}\right)\left\{\sum_j\left[\prod_{k\neq j}\left(\frac{1 + i\lambda_2\omega_k}{1 + i\lambda_1\omega_k}\right)\right]\right\},
\end{equation}
\begin{equation}
    \Omega_{10}(\omega_1,\omega_2,\omega_3;r,s) = \left(\frac{1}{1 + i\lambda_1\underset{j}{\sum}\omega_j}\right)\left[\prod_j\left(\frac{1 + i\lambda_2\omega_j}{1 + i\lambda_1\omega_j}\right)\right].
\end{equation}
\label{eq:cubic_basis}
\end{subequations}
\makebox[\linewidth]{\rule{\textwidth}{0.4pt}}
\end{figure*}
At second and third order with respect to $\gamma_0$, the mathematics become more tedious. For the sake of brevity, we relegate the full derivation to the Supporting Information, and present here only the solutions.

The second order equations indicate that the second order shear stress component $\sigma^{(2)}_{12}$ is zero as expected from symmetry arguments. However, they also provide solutions for the leading order contributions to the normal stress differences in MAPS for the generalized nonlinear Maxwell model, which will be discussed further in Section \ref{sec:normal_stresses}. At third order, we find the analytical solution for the third order complex viscosity, which is presented in equations \ref{eq:cubic_solution} and \ref{eq:cubic_basis} on the following page. 

In equation \ref{eq:cubic_solution}, $\eta^*_3(\omega_1,\omega_2,\omega_3)$ is written as a basis expansion in terms of ten rational functions $\Omega_n(\omega_1,\omega_2,\omega_3;r,s)$, which are given by equation \ref{eq:cubic_basis}. The coefficients $a_n^{(r)}$ and $b_n$ are, in general, nonlinear combinations of the coefficients in the cubic expansion of $F(\boldsymbol{\sigma},\boldsymbol{\dot\gamma})$. Expressions for the coefficients in equation \ref{eq:cubic_solution} are listed in Table \ref{tab:cubic_coefficients}. Equations \ref{eq:cubic_solution} and \ref{eq:cubic_basis} can be used along with the expressions in Table \ref{tab:cubic_coefficients} to find the analytical solution for the third order complex viscosity for all models contained in Table \ref{tab:sub_models}. In this sense, equations \ref{eq:cubic_solution}-\ref{eq:cubic_basis} are a quite compact representation of a diverse set of constitutive models and nonlinear behavior.

\begin{table}[t!]
    \centering
    \begin{tabular}{|C{0.3cm}|C{2.7cm}|C{2.4cm}|C{1.5cm}|} \hline
        $n$ & $a_n^{(0)}$ & $a_n^{(3)}$ & $b_n$ \\[5pt] \hline
        1 & $\lambda_1\lambda_2 - \mu_1(\mu_2 - \nu_2)$ & $\mu_0\left(\mu_2 - \frac{3}{2}\nu_2\right)$ & $2\zeta\frac{\mu_1}{\mu_0}a_1^{(3)}$ \\[5pt] \hline
        2 & $-\lambda_1^2 + \mu_1(\mu_1 - \nu_1)$ & $-\mu_0\left(\mu_1 - \frac{3}{2}\nu_1\right)$ & $2\zeta\frac{\mu_1}{\mu_0}a_2^{(3)}$ \\[5pt] \hline
        3 & $2\alpha_1(\mu_2 - \nu_2)$ & $-2\alpha_0\left(\mu_2 - \frac{3}{2}\nu_2\right)$ & $2\zeta\frac{\alpha_1}{\alpha_0}a_3^{(3)}$ \\[5pt] \hline
        4 & $-2\alpha_1(\mu_1 - \nu_1)$ & $2\alpha_0\left(\mu_1 - \frac{3}{2}\nu_1\right)$ & $2\zeta\frac{\alpha_1}{\alpha_0}a_4^{(3)}$ \\[5pt] \hline
        5 & $-\mu_1(\alpha_1 - \beta_1)$ & $\mu_0\left(\alpha_1 - \frac{3}{2}\beta_1\right)$ & $2\zeta\frac{\mu_1}{\mu_0}a_5^{(3)}$ \\[5pt] \hline
        6 & $2\alpha_1(\alpha_1 - \beta_1)$ & $-2\alpha_0\left(\alpha_1 - \frac{3}{2}\beta_1\right)$ & $2\zeta\frac{\alpha_1}{\alpha_0}a_6^{(3)}$ \\[5pt] \hline
        7 & $2\lambda_1 (d_3 + f_4)$ & 0 & 0 \\[5pt] \hline
        8 & $2\lambda_1(c_2 + d_2 + f_2)$ & 0 & 0 \\[5pt] \hline
        9 & $2\lambda_1(c_1 + d_1 + f_1)$ & 0 & 0 \\[5pt] \hline
        10 & $2\lambda_1 (c_3 + f_3)$ & 0 & 0 \\[5pt] \hline
    \end{tabular}
    \caption{Relationships between the coefficients in the analytical solution for $\eta^*_3(\omega_1,\omega_2,\omega_3)$ (equation \ref{eq:cubic_solution}) to the coefficients in the cubic Maxwell model (equation \ref{eq:cubic_model}).}
    \label{tab:cubic_coefficients}
\end{table}

Despite their ability to represent a very wide range of constitutive viscoelastic models, the expressions for many of the basis functions in equation \ref{eq:cubic_basis} are rather tedious, which might make the task of parameter regression in this model difficult. The parameters $\eta_0$, $\lambda_1$, and $\lambda_2$ can be determined from linear viscoelastic data (equation \ref{eq:eta1}), and MAPS data may be obtained with the goal of determining the coefficients in equation \ref{eq:cubic_solution} as well as the parameter $\zeta$. Although equation \ref{eq:cubic_solution} is nonlinear in $\zeta$, it is linear in all of the lumped coefficients appearing in Table \ref{tab:cubic_coefficients}. Therefore, a suitable protocol for fitting experimental data to equation \ref{eq:cubic_solution} might be to perform linear-least squares fits to the data at different fixed values of $\zeta$, and performing cross-validation on those fits to determine the optimal $\zeta$, similar to how one would choose the optimal value for the regularization hyperparameter in a regularized regression problem.

It is notable that the parameter $\zeta$ has an effect on the third order complex viscosity that is distinct from all other parameters in the cubic Maxwell model. Outside of the parameters that contribute to the linear response ($\eta_0$, $\lambda_1$, and $\lambda_2$), all parameters besides $\zeta$ simply contribute to the coefficients listed in Table \ref{tab:cubic_coefficients}, defining only the scale of different elements in the weakly nonlinear response corresponding to the basis functions in equation \ref{eq:cubic_basis}. The parameter $\zeta$, however, fundamentally changes the behavior of each of these elements, and its effect is likely distinguishable across a wide range of time scales. Out of the models listed in Table \ref{tab:sub_models}, only the versions of the Rolie-Poly and cDCR-CS models for stretching Gaussian chains have nonzero $\zeta$, and in both cases $\zeta$ is related only to the entanglement number, denoted $Z_R$ or $Z_c$ respectively. Thus, properly modeling the rheological consequences of chain entanglements alters the nonlinear response in ways that other physical effects, such as constraint release (associated with $\beta_R$ and $\beta_c$) cannot. Moreover, due to the unique effect of $\zeta$ on the basis functions in equation \ref{eq:cubic_basis}, MAPS provides a convenient framework for observing these entanglement effects.

If the parameter $\zeta$ is set to zero, however, as is the case for most of the models in Table \ref{tab:sub_models}, then we find that the basis functions in equation \ref{eq:cubic_basis} depend only on $\lambda_1$ and $\lambda_2$, and the coefficients $b_n$ all become zero [c.f. Table \ref{tab:cubic_coefficients}], so the third order complex viscosity only depends on the parameters in the polynomial expansion of $F(\boldsymbol{\sigma},\boldsymbol{\dot{\gamma}})$ through the coefficients $a_n^{(0)}$ and $a_n^{(3)}$. This makes the basis functions substantially more compact. In particular, when $\zeta = 0$, we need only consider the basis functions with $r = 0$ and $s = 0$, for which we define:
\begin{equation}
    \Omega_n(\omega_1,\omega_2,\omega_3) \equiv \Omega_n(\omega_1,\omega_2,\omega_3;0,0).
\end{equation}
Moreover, we can further combine the coefficients in the basis expansion for $\eta^*_3(\omega_1,\omega_2,\omega_3)$:
\begin{equation}
    A_n \equiv a_n^{(0)} + a_n^{(3)},
    \label{eq:A_coeff}
\end{equation}
resulting in a simpler basis expansion for the third order complex viscosity, $\eta^*_3(\omega_1,\omega_2,\omega_3)$:
\begin{equation}
    \frac{\eta^*_3(\omega_1,\omega_2,\omega_3)}{\eta_0} = \sum_{n=1}^{N} A_n \Omega_n(\omega_1,\omega_2,\omega_3),
    \label{eq:simpler_solution}
\end{equation}
with $N = 10$. This expansion has the advantageous property that it depends linearly on the unknown coefficients $A_n$. Therefore, given $\eta_0$, $\lambda_1$, and $\lambda_2$ from linear viscoelastic data, regressing the coefficients $A_n$ from MAPS data is a linear problem. For example, given a MAPS data set with $M$ data points at distinct three-frequency coordinates $(\omega_1,\omega_2,\omega_3)_m$, $m = 1\ldots M$, we can construct a $2M \times N$ matrix $\boldsymbol{\Omega}$ whose element $\Omega_{m,n}$, is the value of $\mathrm{Re}[\Omega_n(\omega_1,\omega_2,\omega_3)]$ evaluated at the three-frequency coordinates $(\omega_1,\omega_2,\omega_3)_m$, and whose element $\Omega_{m+M,n}$ is the value of $\mathrm{Im}[\Omega_n(\omega_1,\omega_2,\omega_3)]$ evaluated at the three-frequency coordinates $(\omega_1,\omega_2,\omega_3)_m$. Similarly, we can construct a $2M \times 1$ column vector $\boldsymbol{\eta}_3$ whose $m$th element is the value of
\begin{equation}
    \mathrm{Re}[\eta^*_3(\omega_1,\omega_2,\omega_3)] \equiv \eta'_3(\omega_1,\omega_2,\omega_3)    
\end{equation}
measured at the three-frequency coordinates $(\omega_1,\omega_2,\omega_3)_m$, and whose $(m + M)$th element is the value of
\begin{equation}
    \mathrm{Im}[\eta^*_3(\omega_1,\omega_2,\omega_3)] \equiv -\eta''_3(\omega_1,\omega_2,\omega_3)
\end{equation}
measured at the three-frequency coordinates $(\omega_1,\omega_2,\omega_3)_m$. The coefficients $A_n$ can then be determined straightforwardly by linear least-squares regression:
\begin{equation}
    \textbf{A} = (\boldsymbol{\Omega}^T\boldsymbol{\Omega})^{-1}\boldsymbol{\Omega}^T\boldsymbol{\eta}_3,
    \label{eq:regression_problem}
\end{equation}
where $A_n$ is the $n$th element of the $N\times 1$ column vector $\textbf{A}$.

As previously discussed, many of the models listed in Table \ref{tab:sub_models} fall within the framework of the quadratic Maxwell model presented in equation \ref{eq:quadratic_model}. In the quadratic Maxwell model, $\zeta = 0$; therefore, the third order complex viscosity for this model takes the form of equation \ref{eq:simpler_solution}. Moreover, no cubic terms appear in the quadratic Maxwell model, thus $A_7 = A_8 = A_9 = A_{10} = 0$. Therefore, the third order complex viscosity for the quadratic Maxwell model is equivalent to equation \ref{eq:simpler_solution} with $N = 6$.

Equations \ref{eq:cubic_solution} and \ref{eq:simpler_solution} represent one of this work's primary results, in that they provide, in combination with Table \ref{tab:sub_models}, a library of constitutive model solutions in MAPS rheology. At the same time, they represent the foundation for another problem explored in this work: the problem of quantitative model identification. In this context, model identification may be viewed as the problem of obtaining a suitable parameterization of the underlying constitutive framework in equation \ref{eq:cubic_model}, which may correspond to one of the specific constitutive models in Table \ref{tab:sub_models}. This parameterization may be obtained simply by fitting medium-amplitude data obtained in simple shear to either equation \ref{eq:cubic_solution} or \ref{eq:simpler_solution}, perhaps by using one of the regression protocols discussed previously. In Section \ref{sec:model_identification}, the details of the model identification problem will be explored in greater depth for selected medium-amplitude experiments. For now, though, it is worthwhile to briefly discuss certain important aspects of this problem, especially to identify its limitations.

\subsection{Model Identification Using MAPS Rheology}

Using the tools of MAPS rheology to study the medium-amplitude response of the generalized nonlinear Maxwell model, we have realized one of the principal goals of this work: to obtain a general framework within which quantitative model identification is possible. Once the decision to employ the generalized Maxwell model has been made, model identification involves:
\begin{itemize}
    \item determining $\eta_0$, $\lambda_1$, and $\lambda_2$ from linear viscoelastic data for $\eta^*_1(\omega)$, and
    \item performing parameter regression on medium amplitude data to determine the parameter $\zeta$ and the remaining model parameters using equation \ref{eq:cubic_solution}.
\end{itemize}
We have also shown how one can sacrifice some generality by setting $\zeta = 0$ in order to obtain a simpler regression problem, in the form of equation \ref{eq:regression_problem}.


Regardless of the framework employed for model identification, the MAPS response of the generalized nonlinear Maxwell model presented in Section \ref{sec:maps_solution} reveals some fundamental limitations of MAPS rheology, and any other medium-amplitude simple shear experiments. In particular, the expressions for the coefficients in Table \ref{tab:cubic_coefficients} show that, even if it is possible to accurately determine the coefficients $a_n^{(0)}$, $a_n^{(3)}$, and $b_n$ for the cubic Maxwell model, or the coefficients $A_n$ for the quadratic Maxwell model, it is not possible to distinguish all of the coefficients that appear in equations \ref{eq:cubic_model} and \ref{eq:quadratic_model}. For example, it is clear that the coefficients $c_3$ and $f_3$ can only be inferred as the linear combination $c_3 + f_3$. Therefore, the effects of the cubic terms $(\boldsymbol{\sigma}:\boldsymbol{\sigma})\boldsymbol{\sigma}$ and $\boldsymbol{\sigma}\cdot\boldsymbol{\sigma}\cdot\boldsymbol{\sigma}$, which may be distinct in more highly nonlinear shear deformations or different flow kinematics, are not distinguishable by medium amplitude simple shear measurements. In the quadratic Maxwell model, moreover, there are only six coefficients $A_n$ distinguishable from the MAPS response, which relate to eight of the parameters in equation \ref{eq:quadratic_model}. Thus it is not possible to fully parameterize the quadratic Maxwell model from MAPS measurements of the shear stress. Because each term in the quadratic Maxwell model may produce distinct signatures in highly nonlinear shear stress deformations or other kinematics, we again see that medium amplitude shear stress measurements alone cannot distinguish all nonlinear features of a material's stress-strain relationship.

Though measurements of the linear and weakly nonlinear shear stress response through MAPS rheology evidently cannot fully characterize every aspect of a material's nonlinear response space, shear stress data may not be the only data available in MAPS experiments. In particular, viscoelastic materials undergoing shear deformation produce normal stresses, which may also carry independent information about this response space. In the context of the generalized nonlinear Maxwell model, the normal stress response may provide the remaining information necessary to distinguish all terms in the polynomial expansions of $F(\boldsymbol{\sigma},\boldsymbol{\dot\gamma})$. In the following section, we explore these medium amplitude normal stress effects.

\subsection{Normal Stresses in MAPS Rheology}

The present mathematical framework for MAPS rheology is a scalar description of the functional relationship between the shear stress and shear strain (or strain rate) in time-varying simple shear flow. In general, however, stresses in non-Newtonian fluids are fundamentally tensorial. As Oldroyd notes in his 1958 paper, the ``essential features of observed non-Newtonian behavior'' include ``normal stresses which, in addition to shear stresses, are present in such a liquid in a state of simple shearing flow'' \cite{oldroyd-1958}. These normal stresses contribute to a variety of interesting and industrially important phenomena, such as the Weissenberg effect. Though it has been demonstrated that shear stress measurements using the tools of MAPS rheology provide a detailed characterization of viscoelastic materials, the preceding analysis has revealed that the weakly nonlinear shear stress response is insufficient to distinguish some potentially distinct features of a material's nonlinear response. Because many rheometers that measure the shear stress response of a material to a simple shearing deformation are also capable of measuring some normal stress effects, it is worth exploring what new information this additional independent data may provide in the context of MAPS rheology.

\setcounter{equation}{28}
\begin{figure*}
\begin{align}
    \boldsymbol{\tau}_e(t) = & \int_{-\infty}^{t}M_I(t-t')\boldsymbol{\gamma}(t',t)dt' \nonumber + \iint_{-\infty}^{t}M_{II}(t-t',t-t'')\left(\boldsymbol{\gamma}(t',t)\cdot\boldsymbol{\gamma}(t'',t) + \boldsymbol{\gamma}(t'',t)\cdot\boldsymbol{\gamma}(t',t)\right) \nonumber \\
    & + \iiint_{-\infty}^{t}\left[M_{III}(t-t',t-t'',t-t''')\boldsymbol{\gamma}(t',t)\boldsymbol{\gamma}(t'',t):\boldsymbol{\gamma}(t''',t) \right. \nonumber \\
    & \left. + M_{IV}(t-t',t-t'',t-t''')\left(\boldsymbol{\gamma}(t',t)\cdot\boldsymbol{\gamma}(t'',t)\cdot\boldsymbol{\gamma}(t''',t) + \boldsymbol{\gamma}(t''',t)\cdot\boldsymbol{\gamma}(t'',t)\cdot\boldsymbol{\gamma}(t',t)\right)\right]dt'dt''dt''' + O(\boldsymbol{\gamma}^4). \label{eq:memory_integral}
\end{align}
\makebox[\linewidth]{\rule{\textwidth}{0.4pt}}
\end{figure*}
\setcounter{equation}{24}

Experimentally, the normal stresses that exist within a sample can be measured only relative to an isotropic contribution. Thus, rheologists typically study normal stress differences. The first normal stress difference in simple shear, $\sigma_{11}(t) - \sigma_{22}(t)$, is typically denoted by $N_1(t)$ and the second normal stress difference, $\sigma_{22}(t) - \sigma_{33}(t)$, is denoted by $N_2(t)$ \cite{nomenclature}. The mathematical framework for MAPS rheology may be extended to include descriptions of these normal stress differences by defining two additional Volterra series expansions:
\begin{align}
    \hat{N}_1(\omega) =& \sum_{n\in \mathrm{even}}\frac{1}{(2\pi)^{n-1}}\int\ldots\int_{-\infty}^{\infty}\Psi^{*(n)}_{1}(\omega_1,...,\omega_n) \nonumber \\
    & \times \delta(\omega - \sum_{m=1}^n\omega_m) \prod_{m=1}^n \hat{\dot\gamma}(\omega_m)d\omega_m,
\end{align}
\begin{align}
    \hat{N}_2(\omega) =& \sum_{n\in \mathrm{even}}\frac{1}{(2\pi)^{n-1}}\int\ldots\int_{-\infty}^{\infty}\Psi^{*(n)}_{2}(\omega_1,...,\omega_n) \nonumber \\
    & \times \delta(\omega - \sum_{m=1}^n\omega_m) \prod_{m=1}^n \hat{\dot\gamma}(\omega_m)d\omega_m,
\end{align}
which have been written in terms of the Fourier transforms of the time-varying normal stress differences and the shear strain rate. These expressions are analogous to the Volterra series expansion of the shear stress in terms of the strain rate, though they now reflect the necessarily even symmetry of the normal stresses. Thus we call the response function $\Psi_1^{*(n)}$ (or $\Psi_2^{*(n)}$) the \emph{nth order first (or second) normal stress coefficient}. In MAPS rheology, only the leading order nonlinear term in the Volterra series is considered. For normal stresses, which are inherently nonlinear phenomena in simple shear, this is equivalent to the leading order term in the Volterra series, which appears at second order. In MAPS rheology, therefore, expressions for the first and second normal stress differences are:
\begin{align}
    \hat{N}_1(\omega) =& \frac{1}{2\pi}\iint_{-\infty}^{\infty}\tilde\Psi^{*}_{1}(\omega_1,\omega_2)\delta(\omega - \omega_1 - \omega_2) \nonumber \\
    & \times \hat{\dot\gamma}(\omega_1)\hat{\dot\gamma}(\omega_2)d\omega_1 d\omega_2 + O(\hat{\dot\gamma}^4),
    \label{eq:maps_N1}
\end{align}
\begin{align}
    \hat{N}_2(\omega) =& \frac{1}{2\pi}\iint_{-\infty}^{\infty}\tilde\Psi^{*}_{2}(\omega_1,\omega_2)\delta(\omega - \omega_1 - \omega_2) \nonumber \\
    & \times \hat{\dot\gamma}(\omega_1)\hat{\dot\gamma}(\omega_2)d\omega_1 d\omega_2 + O(\hat{\dot\gamma}^4),
    \label{eq:maps_N2}
\end{align}
where we call $\tilde\Psi^*_1(\omega_1,\omega_2) \equiv \Psi^{*(2)}_1(\omega_1,\omega_2)$ the \emph{leading order first normal stress coefficient}, and $\tilde\Psi^*_2(\omega_1,\omega_2) \equiv$ $\Psi^{*(2)}_2(\omega_1,\omega_2)$ the \emph{leading order second normal stress coefficient}.

Equations \ref{eq:maps_N1} and \ref{eq:maps_N2} along with equation \ref{eq:MAPS} form the most general tensorial description of the stress-strain relationship in MAPS rheology. The four response functions $\tilde\Psi^*_1(\omega_1,\omega_2)$, $\tilde\Psi^*_2(\omega_1,\omega_2)$, $\eta^*_1(\omega)$, and $\eta^*_3(\omega_1,\omega_2,\omega_3)$ that govern these expressions are not necessarily related to each other in this most general case. However, many viscoelastic fluids and constitutive models -- including, by construction, the cubic Maxwell model -- fall into a more specific category: materials that are isotropic. For these materials and models, the response functions that govern the shear stress and normal stress responses in MAPS rheology are not independent, but rather are connected by a general, tensorial polynomial expansion called the \emph{memory integral expansion}. In the following section, we will briefly introduce this expansion and demonstrate how it connects the response functions that govern this new tensorial description of MAPS rheology.

\subsubsection{The Memory Integral Expansion in Simple Shear}

In writing the scalar Volterra series expansion of shear stress in terms of shear strain, we have made the underlying assumption that the shear stress is described exactly by some nonlinear, time-invariant functional of the shear strain. For isotropic materials, we can make the broader statement that the extra stress tensor is a nonlinear, time-invariant functional of some suitable frame-invariant tensor describing the accumulated deformation in a material. One commonly employed choice for this tensorial measure of the deformation is the relative finite strain tensor, denoted here by $\boldsymbol{\gamma}(t,t')$, which measures the deformation accumulated between the present time $t$ and a time $t'$ in the past. Such a functional relationship should still possess a polynomial approximation, which we may write using the same theory of matrix polynomials used to develop the cubic Maxwell model \cite{spencer-1958}. Using the simplifications proposed by Pipkin, this polynomial approximation reduces to the memory integral expansion \cite{pipkin-1964}, which is presented in equation \ref{eq:memory_integral}.

To third order in the relative finite strain tensor, the memory integral expansion is described by only four response functions, $M_I$, $M_{II}$, $M_{III}$, and $M_{IV}$. This echoes the structure of the tensorial description of MAPS rheology, which also includes only four independent response functions.

In simple shear, we are able to relate the MAPS response functions to the response functions in the memory integral expansion. To do so, we first note that the relative finite strain tensor in simple shear takes the form:
\setcounter{equation}{29}
\begin{equation}
    \boldsymbol{\gamma}(t',t) = \begin{pmatrix} \gamma^2(t',t) & \gamma(t',t) & 0 \\
    \gamma(t',t) & 0 & 0 \\
    0 & 0 & 0 \end{pmatrix},
\end{equation}
with the accumulated strain measure:
\begin{equation}
    \gamma(t',t) = \int_{t'}^t \dot\gamma(t'')dt''.
\end{equation}
At first order in the shear rate $\dot\gamma$, the relationship between $\eta^*_1(\omega)$ and $M_{I}(t)$ emerges:
\begin{equation}
    \eta^*_1(\omega) = -\frac{1}{i\omega}[M^*_{I}(\omega) - M^*_{I}(0)],
\end{equation}
which has been compactly represented in terms of the Fourier transform of $M_{I}(t)$:
\begin{equation}
    M^*_{I}(\omega) = \int_{-\infty}^{\infty}M_{I}(t)e^{-i\omega t}dt.
\end{equation}
The relationship between the first memory kernel $M_{I}(t-t')$ and familiar linear viscoelastic properties is well-known, though typically represented in terms of $M_{I}(t-t')$ and the linear relaxation modulus $G(t-t')$: $M_{I}(t-t') = \partial G(t-t')/\partial t'$ \cite{martinetti-2019,vrentas-1991}.

At second order in the shear rate, we find relationships between $\tilde\Psi^*_1(\omega_1,\omega_2)$, $\tilde\Psi^*_2(\omega_1,\omega_2)$, $M_{I}(t)$, and $M_{II}(t_1,t_2)$. Interestingly, $\tilde\Psi^*_1(\omega_1,\omega_2)$ relates only to the linear response function, $M_{I}(t)$, and does so in a manner such that it is possible to express $\tilde\Psi^*_1(\omega_1,\omega_2)$ entirely in terms of $\eta^*_1(\omega)$:
\begin{align}
    \bar{\Psi}^*_1(\omega_1,\omega_2) = & \frac{i}{\omega_1 \omega_2}\left[ (\omega_1 + \omega_2)\eta^*_1(\omega_1 + \omega_2) \right. \nonumber \\
    & \quad \quad \quad \left. - \omega_1\eta^*_1(\omega_1) - \omega_2\eta^*_1(\omega_2)\right].
    \label{eq:psi1_eta1}
\end{align}
In simple shear rheology, the leading order first normal stress difference provides no more information about an isotropic simple fluid than does the linear shear stress response. While this result is known for steady and single-tone oscillatory flows \cite{vrentas-1991}, equation \ref{eq:psi1_eta1} extends this conclusion to all simple shear flows. Unfortunately, the first normal stress difference is often the easiest normal stress difference to measure experimentally. Therefore, in practice, nothing about an isotropic material is revealed by first normal stress differences in medium-amplitude experiments that is not already evident from the shear stress response. In the context of the cubic and quadratic Maxwell models, this means that those parameters that could only be identified in lumped coefficients by the third-order shear stress response cannot be segregated by the observed leading order contribution to the first normal stress difference. However, because the memory integral expansion is valid only for isotropic materials, any inconsistency between the leading order contribution to the first normal stress difference and the linear shear stress response is a useful indicator that a material is not deforming isotropically in simple shear.

Further examination of the memory integral expansion reveals that the leading order second normal stress coefficient relates only to the response function $M_{II}(t_1,t_2)$, and that the third order complex viscosity relates to a specific linear combination of $M_{II}(t_1,t_2)$, $M_{III}(t_1,t_2,t_3)$, and $M_{IV}(t_1,t_2,t_3)$. The specific mathematical expressions for these relationships are difficult to express compactly, and are not important for the present analysis; therefore, we present them in the Supporting Information. Still, we see that, in principle, a material's observed second normal stress difference in a MAPS experiment may provide distinct information from that contained within its shear stress response. This observation will be exemplified shortly in the analytical solution to the generalized Maxwell model.

It is important to note that the third order complex viscosity only provides information about a specific linear combination of the two third-order functions in the memory integral expansion: $M_{III}(t_1,t_2,t_3)$ and $M_{IV}(t_1,t_2,t_3)$. For other flow kinematics, such as shear-free flows, these response functions may combine in different proportions or ways than in simple shear. Therefore, even if it is possible to fully characterize $\eta^*_1(\omega)$, $\eta^*_3(\omega_1,\omega_2,\omega_3)$, $\tilde\Psi^*_1(\omega_1,\omega_2)$, and $\tilde\Psi^*_2(\omega_1,\omega_2)$, this is not necessarily sufficient to predict the medium amplitude response to different kinematic histories. Further exploration of the weakly nonlinear signatures to flows of different kinematics, and their relationship to the MAPS response functions, is left to future studies. \newline

\subsubsection{Leading Order Normal Stress Response of the Generalized Nonlinear Maxwell Model}
\label{sec:normal_stresses}

As discussed in Section \ref{sec:maps_solution}, the process of obtaining the solution for the third order complex viscosity for the generalized Maxwell model by asymptotic analysis produces a set of differential equations at second order in $\gamma_0$ that describe the leading order normal stress contributions. These differential equations may be manipulated to obtain analytical expressions for the leading order first and second normal stress coefficients. These differential equations, as well as the mathematical steps to obtain the analytical forms for $\tilde\Psi^*_1(\omega_1,\omega_2)$ and $\tilde\Psi^*_2(\omega_1,\omega_2)$, are presented in the Supporting Information for the sake of brevity. The resulting leading order first and second normal stress differences for the generalized Maxwell model are compactly represented as basis expansions:
\begin{equation}
    \frac{\tilde\Psi^*_1(\omega_1,\omega_2)}{\eta_0} = \sum_{n=1}^{3}B_n \psi_n(\omega_1,\omega_2),
    \label{eq:psi1}
\end{equation}
\begin{equation}
    \frac{\tilde\Psi^*_2(\omega_1,\omega_2)}{\eta_0} = \sum_{n=1}^{3}C_n \psi_n(\omega_1,\omega_2),
    \label{eq:psi2}
\end{equation}
where the functions $\psi_n(\omega_1,\omega_2)$ are:
\begin{equation}
    \psi_1(\omega_1,\omega_2) = \left(\frac{1}{1 + i\lambda_1\underset{j=1}{\overset{2}{\sum}}\omega_j}\right),
    \label{eq:ns_basis1}
\end{equation}
\begin{equation}
    \psi_2(\omega_1,\omega_2) = \frac{1}{2}\left(\frac{1}{1 + i\lambda_1\underset{j=1}{\overset{2}{\sum}}\omega_j}\right) \left[\sum_{j=1}^2\left(\frac{1 + i\lambda_2\omega_j}{1 + i\lambda_1\omega_j},\right)\right],
\end{equation}
\begin{equation}
    \psi_3(\omega_1,\omega_2) = \left(\frac{1}{1 + i\lambda_1\underset{j=1}{\overset{2}{\sum}}\omega_j}\right) \left[\prod_{j=1}^2\left(\frac{1 + i\lambda_2\omega_j}{1 + i\lambda_1\omega_j}\right)\right].
\end{equation}
The coefficients $B_n$ and $C_n$ are related to some of the coefficients in equation \ref{eq:cubic_model}:
\begin{equation}
    B_1 = -2\lambda_2, \quad B_2 = 2\lambda_1, \quad B_3 = 0,
    \label{eq:B_coeff}
\end{equation}
and
\begin{equation}
    C_1 = \lambda_2 - \mu_2, \quad C_2 = -(\lambda_1 - \mu_1), \quad C_3 = -\alpha_1.
    \label{eq:C_coeff}
\end{equation}

Equations \ref{eq:psi1} through \ref{eq:C_coeff} represent the full solution for the leading order first and second normal stress coefficients for the generalized nonlinear Maxwell model, expressed in terms of the coefficients in the cubic expansion of the model in equation \ref{eq:cubic_model}. Notably, the coefficients in equations \ref{eq:B_coeff} and \ref{eq:C_coeff} do not depend on the parameter $\zeta$, which contributes in an isotropic manner to the normal stresses in the cubic Maxwell model, or on any parameter associated with a cubic term in the model. Therefore, equations \ref{eq:psi1} through \ref{eq:C_coeff} also represent the solution for the quadratic Maxwell model in equation \ref{eq:quadratic_model}. Furthermore, by setting $\alpha_1 = 0$ and applying equations \ref{eq:psi1} through \ref{eq:C_coeff} to a single-tone oscillatory shear protocol in equations \ref{eq:maps_N1} and \ref{eq:maps_N2}, one can validate that these solutions reduce exactly to those previously obtained for the Oldroyd 8-constant model in MAOS \cite{booij-1966,saengow-2017-2}. Specifically, the values of $\tilde{\Psi}^*_1(\omega,\omega)$ and $\tilde{\Psi}^*_2(\omega,\omega)$ dictate the second harmonic response, and the values of $\tilde{\Psi}^*_1(\omega,-\omega)$ and $\tilde{\Psi}^*_2(\omega,-\omega)$ set the constant offset from zero observed in these experiments.

The coefficients $B_n$ that dictate the form of the leading order first normal stress coefficient depend only on $\lambda_1$ and $\lambda_2$, which are both parameters that affect the linear shear stress response. This is expected for isotropic models such as the generalized nonlinear Maxwell model and its polynomial expansions. Indeed, substituting the linear complex viscosity for the model (equation \ref{eq:eta1}) into the expected relationship between $\eta^*_1(\omega)$ and $\tilde\Psi^*_1(\omega_1,\omega_2)$ (equation \ref{eq:psi1_eta1}) recovers the solution defined by equations \ref{eq:psi1} and \ref{eq:ns_basis1} through \ref{eq:B_coeff}.

Unlike the leading order first normal stress coefficient, the leading order second normal stress coefficient does relate to the nonlinear parameters in the model. Interestingly, with $\lambda_1$ and $\lambda_2$ known from linear viscoelastic data, the coefficients $C_n$ allow $\mu_1$, $\mu_2$, and $\alpha_1$ to be uniquely determined. This is indeed distinct from the information provided by the third order complex viscosity, from which $\mu_1$, $\mu_2$, and $\alpha_1$ can only be determined in combination with other model parameters. Therefore, if second normal stress data is available, it is possible to determine all parameters that appear in the quadratic Maxwell model. In principle, therefore, second normal stress data can be quite useful, though in practice it may be difficult to obtain.

\section{Comment on Time-Strain Separability}
\label{sec:time_strain_separability}

A particularly interesting phenomenon in nonlinear rheology is that of time-strain separability (TSS). The principle of TSS is that the nonlinear rheological response of a material can be separated into a time-dependent component -- the linear relaxation modulus $G(t-t')$ -- and a strain-dependent component, often called the damping function $h(\gamma(t,t'))$ \cite{rolon-2009}. In medium amplitude simple shear rheology, TSS dictates that the third-order material response should be specified up to a single scaling factor by the linear viscoelastic response. This has been expressed in terms of mathematical conditions for intrinsic nonlinear functions that are necessary, but not sufficient, for a material to be considered TSS \cite{martinetti-2019}. In MAPS rheology, the condition for TSS is \cite{lennon-2020-1}:
\begin{align}
& G_3^*( \omega_1, \omega_2, \omega_3 ) = \left. \frac{\partial h}{\partial \gamma^2} \right|_{\gamma = 0} \left[ G_1^*\left( \sum_{j=1}^3 \omega_j \right) \right. \label{eq:tss} \\
& \left. \quad  - \sum_{j=1}^3 G_1^*\left( \sum_{\substack{k=1 \\ k\ne j}}^3 \omega_k \right) + \sum_{j=1}^3  G_1^*( \omega_j ) - G_1^*( 0 ) \right]. \nonumber
\end{align}
which is expressed in terms of the leading order derivative of the damping function, and the first and third order complex moduli. The first and third order complex moduli may be computed directly from the first and third order complex viscosities:
\begin{equation}
    G^*_n(\omega_1,\ldots,\omega_n) = \left(\prod_{m=1}^{n} i\omega_m \right) \eta^*_n(\omega_1,\ldots,\omega_n).
\end{equation}
Other authors have used signatures in medium amplitude oscillatory shear (MAOS) to classify constitutive models as either non-TSS or consistent with TSS \cite{yong-2020}. Equation \ref{eq:tss} imposes a slightly stricter condition, in that it is valid for arbitrary medium amplitude, simple shear deformations, rather than just for single-tone oscillations. However, the authors are not aware of any constitutive model that exhibits TSS signatures in MAOS but not in another medium amplitude shear deformation.

Substituting the linear response of the generalized nonlinear Maxwell model (equation \ref{eq:eta1}) into equation \ref{eq:tss} reveals a very simple condition for TSS within this framework:
\begin{equation}
    \frac{a_1^{(0)} + a_1^{(3)}}{a_2^{(0)} + a_2^{(3)}} = \frac{A_1}{A_2} = -\frac{\lambda_2}{\lambda_1},
\end{equation}
with $\zeta$ and all other coefficients in Table \ref{tab:cubic_coefficients} equalling zero. This is a necessary, but not sufficient, condition for any model within the framework of the generalized nonlinear Maxwell model to be considered TSS. These conditions can be applied to all models within Table \ref{tab:sub_models} to classify the models as either non-TSS or consistent with TSS to third order in simple shear. The result of this classification is presented in Table \ref{tab:tss}. This classification is in agreement with that found in a recent study by Hyun and co-workers, with the notable exception that the Larson model was therein classified as TSS, while we have classified it as non-TSS \cite{yong-2020}. However, other authors have confirmed that the Larson model is indeed non-TSS, in agreement with our results \cite{ramlawi-2020}.

\begin{table}[t!]
    \centering
    \begin{tabular}{|C{4cm}|C{4cm}|} \hline
        Consistent with TSS & Non-TSS \\\hline\hline
        Gordon-Schowalter & Non-stretching Gaussian Rolie-Poly \\
        Johnson-Segalman & Stretching Gaussian Rolie-Poly \\
        Oldroyd Fluid A & Stretching Gaussian cDCR-CS \\
        Oldroyd Fluid B & Larson \\
        Arbitrary Normal Stress Ratio (ANSR) & Oldroyd 8-Constant \\
        Corotational Jeffreys & Oldroyd 6-Constant \\
        Williams 3-Constant & Oldroyd 4-Constant \\
        Corotational Maxwell & Second Order Fluid \\
        Upper Convected Maxwell & Denn Modified Maxwell \\
        Lower Convected Maxwell & Giesekus \\
         & Linearized Phan-Thien--Tanner \\\hline
    \end{tabular}
    \caption{The classification of all models in Table \ref{tab:sub_models} as either consistent with TSS to third order in simple shear or non-TSS based on the condition given by equation \ref{eq:tss}.}
    \label{tab:tss}
\end{table}

With some algebra, the expression for the third order complex viscosity for TSS models with a single-mode Jeffreys linear response can be written in another form:
\begin{align}
    \frac{\eta^*_3(\omega_1,\omega_2,\omega_3)}{\eta_0} &= -\frac{\lambda_2}{\lambda_1}A_2\Omega_1(\omega_1,\omega_2,\omega_3) + A_2\Omega_2(\omega_1\omega_2,\omega_3) \nonumber \\
    &= \left(1 - \frac{\lambda_2}{\lambda_1}\right)A_2 \Omega_2(\omega_1,\omega_2,\omega_3; \lambda_2 = 0).
    \label{eq:tss_simplified}
\end{align}
That is, the third order complex viscosity can be rewritten with $\lambda_2$ as a front factor only. That this manipulation is possible is perhaps unsurprising, as it is simply a statement that the retardation term associated with $\lambda_2$ represents a Newtonian solvent mode in the generalized nonlinear Maxwell model. It is of course possible to instead write equation \ref{eq:generalized_nonlinear_maxwell} without this term, then add a Newtonian solvent mode to the polymeric stress governed by the model. The solvent mode does not contribute to the nonlinear response, therefore the TSS expression for the third order complex viscosity would include only the TSS signature of equation \ref{eq:generalized_nonlinear_maxwell} with $\lambda_2 = 0$. The front factor in the second equality of equation \ref{eq:tss_simplified} arises from the relationship between the zero shear viscosity in the single mode model that contains a retardation term, and the zero shear viscosity in the polymeric mode of the model with separate polymeric and solvent modes.

This reformulation is also useful from the perspective of model identification, in that a nonzero value of $\lambda_2$ renders the inversion of equation \ref{eq:regression_problem} more poorly conditioned than for $\lambda_2 = 0$. In fact, when $\lambda_2 = \lambda_1$, many of the basis functions are no longer distinguishable from one another, thus equation \ref{eq:regression_problem} is no longer invertible. Therefore, in the following sections, we will consider only the case where $\lambda_2 = 0$ unless otherwise specified. In any case, the effects of a nonzero $\lambda_2$ can be recovered simply by adding a Newtonian solvent mode to the polymeric mode governed by the generalized nonlinear Maxwell model.

\section{MAPS Rheology Applied to Simple Shear Flows}
\label{sec:applied}

The previous analysis has used the tools of MAPS rheology to elucidate the medium amplitude response of the generalized Maxwell model to simple shear deformations, without making reference to any specific shear deformation protocol. In principle, the data from any simple shear deformation protocol could be used in the model identification problem; however, some data may be more useful than others. In this section, we will apply the MAPS response of the generalized nonlinear Maxwell model to obtain analytical solutions for the model in a few common simple shear deformation protocols, and these results will be used to inform a quantitative exploration of model identification in Section \ref{sec:model_identification}.

\subsection{Three-Tone Oscillatory Shear}
\label{sec:three_tones}

\begin{figure}
    \centering
    \includegraphics[width=0.8\columnwidth]{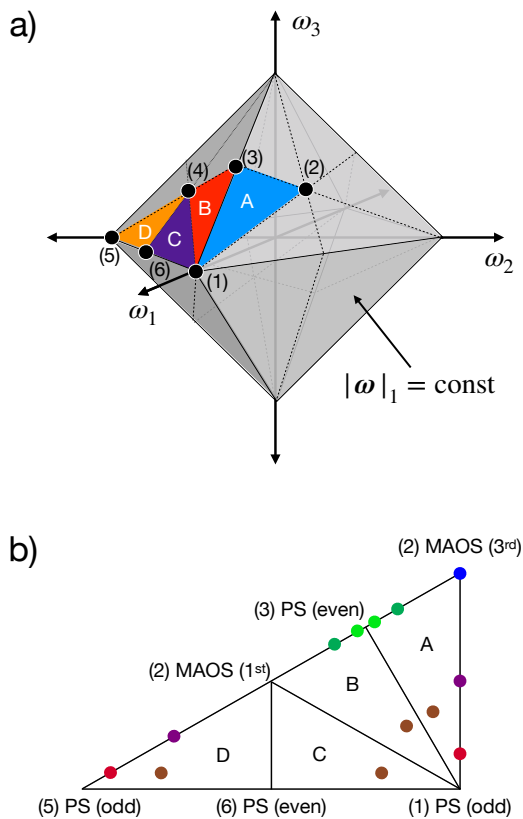}
    \caption{Geometric representation of the three-frequency MAPS domain: $(\omega_1,\omega_2,\omega_3)$. a) The octahedral surface representing a constant $L^1$-norm ($|\boldsymbol{\omega}|_1 = |\omega_1| + |\omega_2| + |\omega_3|$) manifold. Symmetries of the MAPS response function reduce the domain within which the response function takes on unique values to four triangular subspaces, which are labelled A, B, C, and D. b) The projection of the four MAPS subspaces onto the plane, depicting the barycentric coordinates studied by a three-tone experiment with $\{n_1,n_2,n_3\} = \{1,4,16\}$. Points are colored by associating their barycentric coordinates $(r,g,b)$ with RGB color channels.}
    \label{fig:plotting}
\end{figure}

In recent work, it was demonstrated that a shear deformation protocol composed of three sinusoidal oscillations at different frequencies imposed in parallel provides an effective means for broadly studying the three-dimensional domain $(\omega_1,\omega_2,\omega_3)$ of the third order complex viscosity with high data throughput \cite{lennon-2020-2}. Measuring the weakly nonlinear stress response to a single three-tone oscillatory signal of the form:
\begin{equation}
    \gamma(t) = \gamma_0 \sum_j \sin(n_j\omega_0 t)
\end{equation}
can reveal the value of the third order complex viscosity at 19 distinct points within its domain. Separating the real and imaginary components of the response function produces a large data set of 38 independent values. The integers $n_j \in \mathbb{Z}$ set the locations on constant $L^1$-norm surfaces in three-frequency space studied by the experiment, irrespective of the frequency scale of this surface, which is set by the fundamental frequency $\omega_0$. Figure \ref{fig:plotting}a depicts such a constant $L^1$-norm surface, which in three dimensions is equivalent to the surface of a regular octahedron. 

Due to certain symmetries of the MAPS response functions, only coordinates that fall within one of four identical hemi-equilateral triangular subspaces represent distinct values of the response functions. It can be shown that the vertices of these triangular subspaces relate to features observed in single-tone MAOS experiments and parallel superposition experiments, where a single-oscillatory tone is superimposed upon steady shear flow. The four subspaces are distinguished in Figure \ref{fig:plotting}a, and they are removed from the octahedral surface and laid flat in the plane in Figure \ref{fig:plotting}b. Within each of these triangular subspaces we can establish a barycentric coordinate system $(r,g,b)$, and these coordinates can be used as color channels for distinguishing coordinates on shared $L^1$-norm surfaces. For instance, the three-tone experiment with $\{n_1,n_2,n_3\} = \{1,4,16\}$ produces the distribution of coordinates depicted in \ref{fig:plotting}b, regardless of the fundamental frequency $\omega_0$ of the experiment. By separating the frequency-scale and coordinate distribution features of the experimental design, it is possible to conduct three-tone frequency sweeps by varying $\omega_0$ with a constant set of $n_j$. These sweeps can be visualized using familiar tools such as Bode or Nyquist diagrams, with the barycentric color channels used to denote different locations on the constant $L^1$-norm surfaces. For more detail on the three-tone experimental protocol and on methods for visualizing MAPS data sets, we refer readers to the detailed development of MAPS rheology in \cite{lennon-2020-1,lennon-2020-2}.

\begin{figure*}
    \centering
    \includegraphics[width=0.9\textwidth]{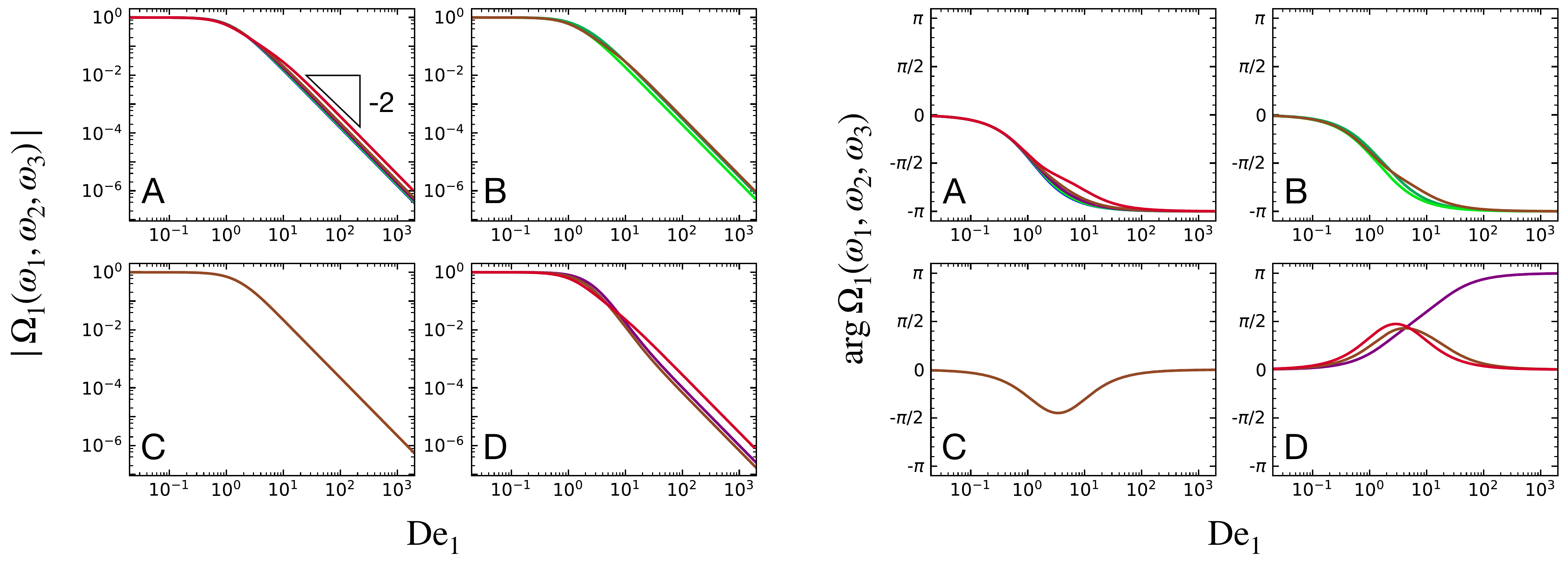}
    \caption{The magnitude (left) and phase (right) of the function $\Omega_1(\omega_1,\omega_2,\omega_3)$ with $\lambda_2 = 0$ along the manifolds studied by a three-tone frequency sweep with $\{n_1,n_2,n_3\} = \{1,4,16\}$, separated into subspaces A, B, C, and D within the MAPS domain. Colors distinguish different barycentric coordinates on the constant $L^1$-norm projection of the MAPS domain, corresponding to the locations in Figure \ref{fig:plotting}b.}
    \label{fig:bode_omega1}
\end{figure*}
\begin{figure*}
    \centering
    \includegraphics[width=0.9\textwidth]{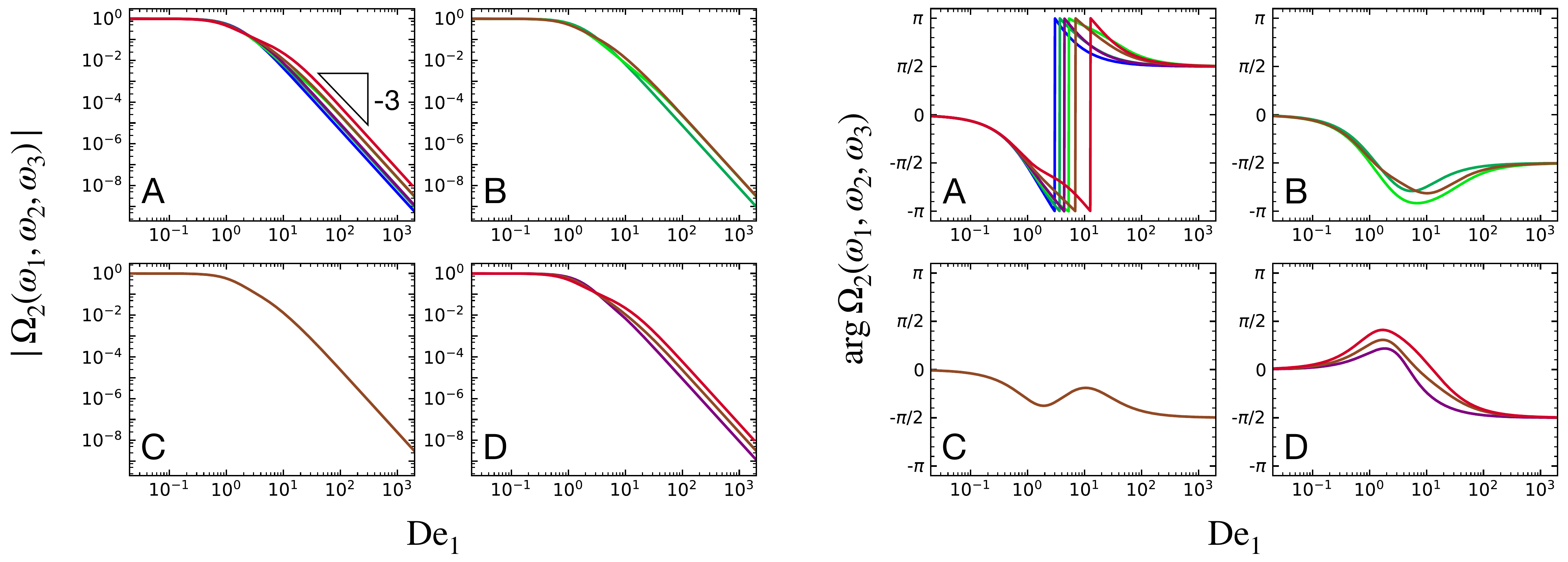}
    \caption{The magnitude (left) and phase (right) of the function $\Omega_2(\omega_1,\omega_2,\omega_3)$ with $\lambda_2 = 0$ along the manifolds studied by a three-tone frequency sweep with $\{n_1,n_2,n_3\} = \{1,4,16\}$, separated into subspaces A, B, C, and D within the MAPS domain. Colors distinguish different barycentric coordinates on the constant $L^1$-norm projection of the MAPS domain, corresponding to the locations in Figure \ref{fig:plotting}b.}
    \label{fig:bode_omega2}
\end{figure*}
\begin{figure*}
    \centering
    \includegraphics[width=0.9\textwidth]{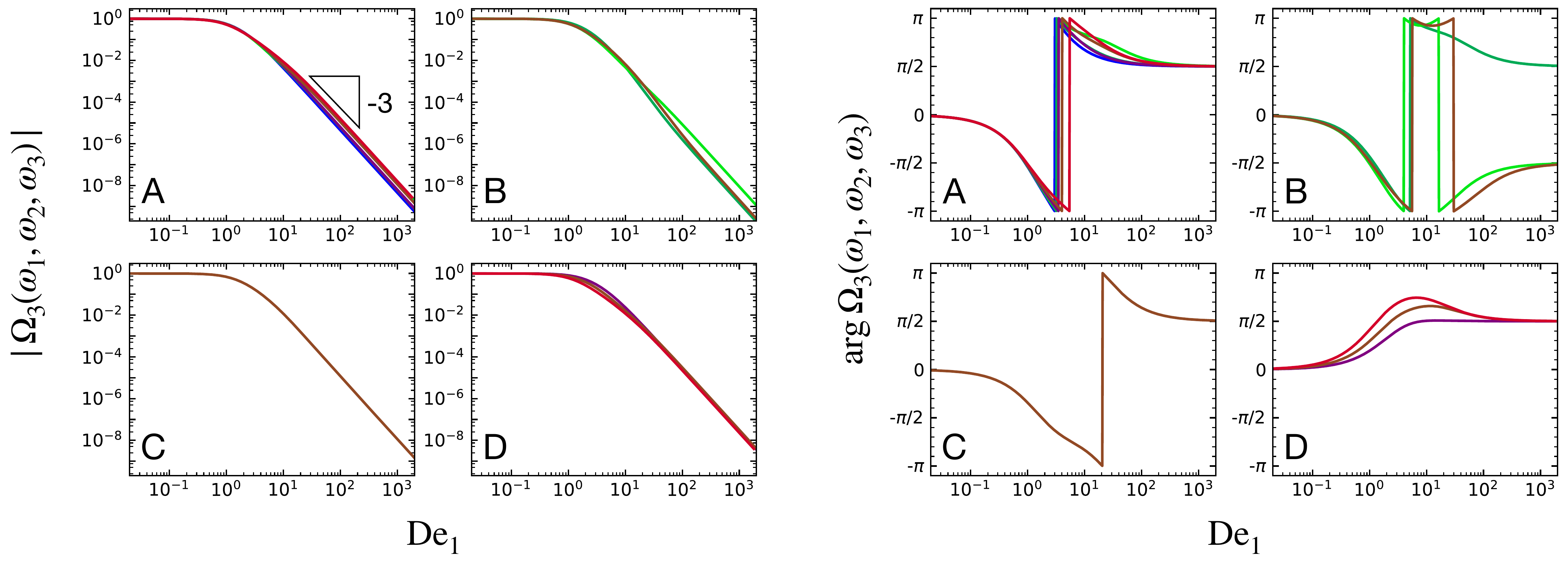}
    \caption{The magnitude (left) and phase (right) of the function $\Omega_3(\omega_1,\omega_2,\omega_3)$ with $\lambda_2 = 0$ along the manifolds studied by a three-tone frequency sweep with $\{n_1,n_2,n_3\} = \{1,4,16\}$, separated into subspaces A, B, C, and D within the MAPS domain. Colors distinguish different barycentric coordinates on the constant $L^1$-norm projection of the MAPS domain, corresponding to the locations in Figure \ref{fig:plotting}b.}
    \label{fig:bode_omega3}
\end{figure*}
\begin{figure*}
    \centering
    \includegraphics[width=0.9\textwidth]{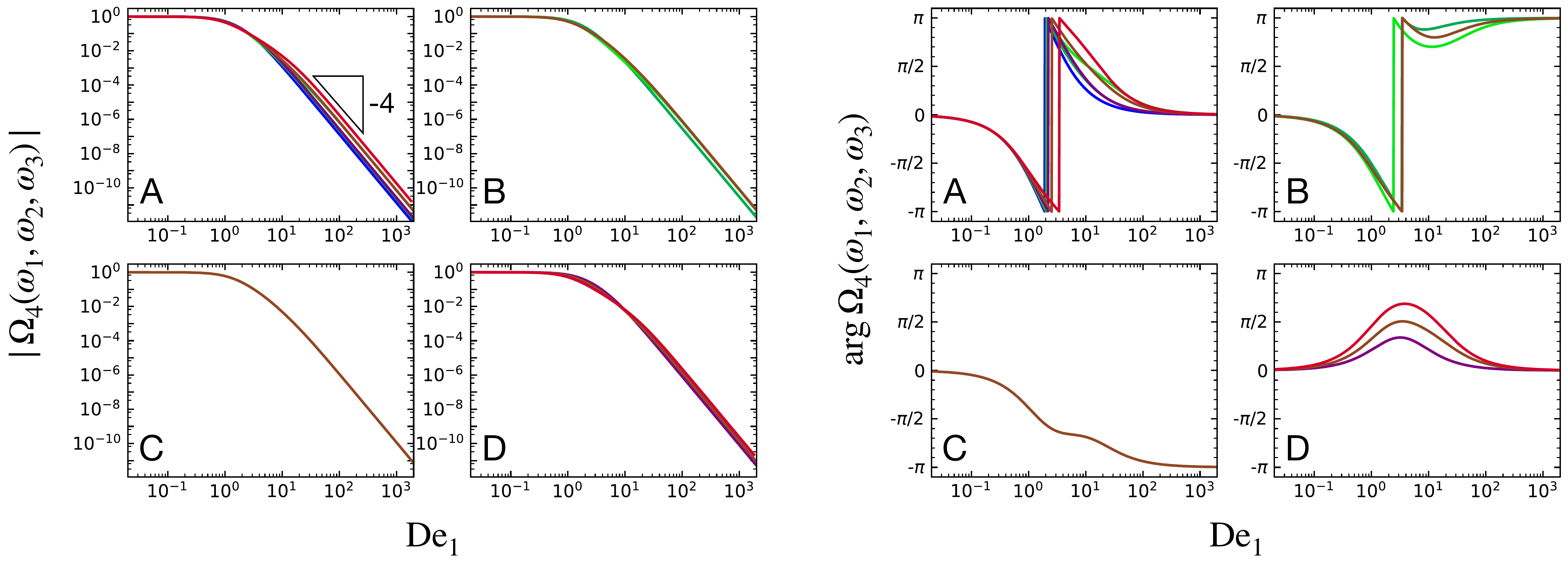}
    \caption{The magnitude (left) and phase (right) of the function $\Omega_4(\omega_1,\omega_2,\omega_3)$ with $\lambda_2 = 0$ along the manifolds studied by a three-tone frequency sweep with $\{n_1,n_2,n_3\} = \{1,4,16\}$, separated into subspaces A, B, C, and D within the MAPS domain. Colors distinguish different barycentric coordinates on the constant $L^1$-norm projection of the MAPS domain, corresponding to the locations in Figure \ref{fig:plotting}b.}
    \label{fig:bode_omega4}
\end{figure*}
\begin{figure*}
    \centering
    \includegraphics[width=0.9\textwidth]{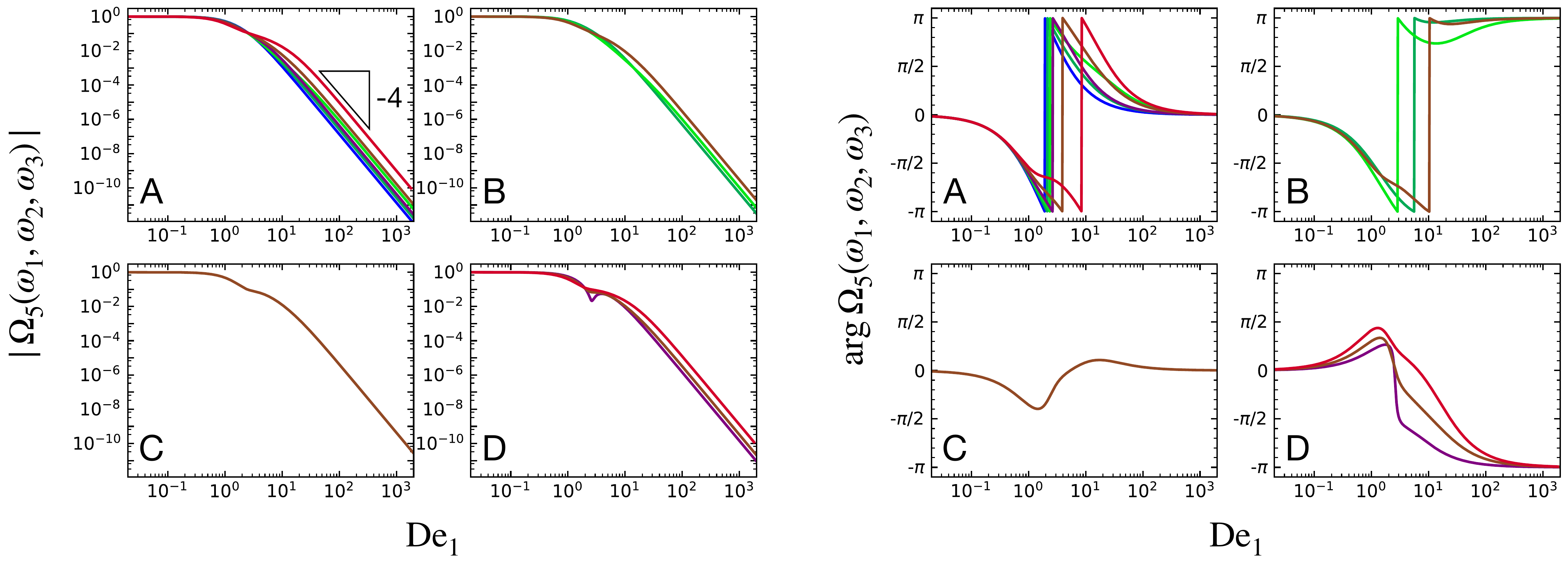}
    \caption{The magnitude (left) and phase (right) of the function $\Omega_5(\omega_1,\omega_2,\omega_3)$ with $\lambda_2 = 0$ along the manifolds studied by a three-tone frequency sweep with $\{n_1,n_2,n_3\} = \{1,4,16\}$, separated into subspaces A, B, C, and D within the MAPS domain. Colors distinguish different barycentric coordinates on the constant $L^1$-norm projection of the MAPS domain, corresponding to the locations in Figure \ref{fig:plotting}b.}
    \label{fig:bode_omega5}
\end{figure*}
\begin{figure*}
    \centering
    \includegraphics[width=0.9\textwidth]{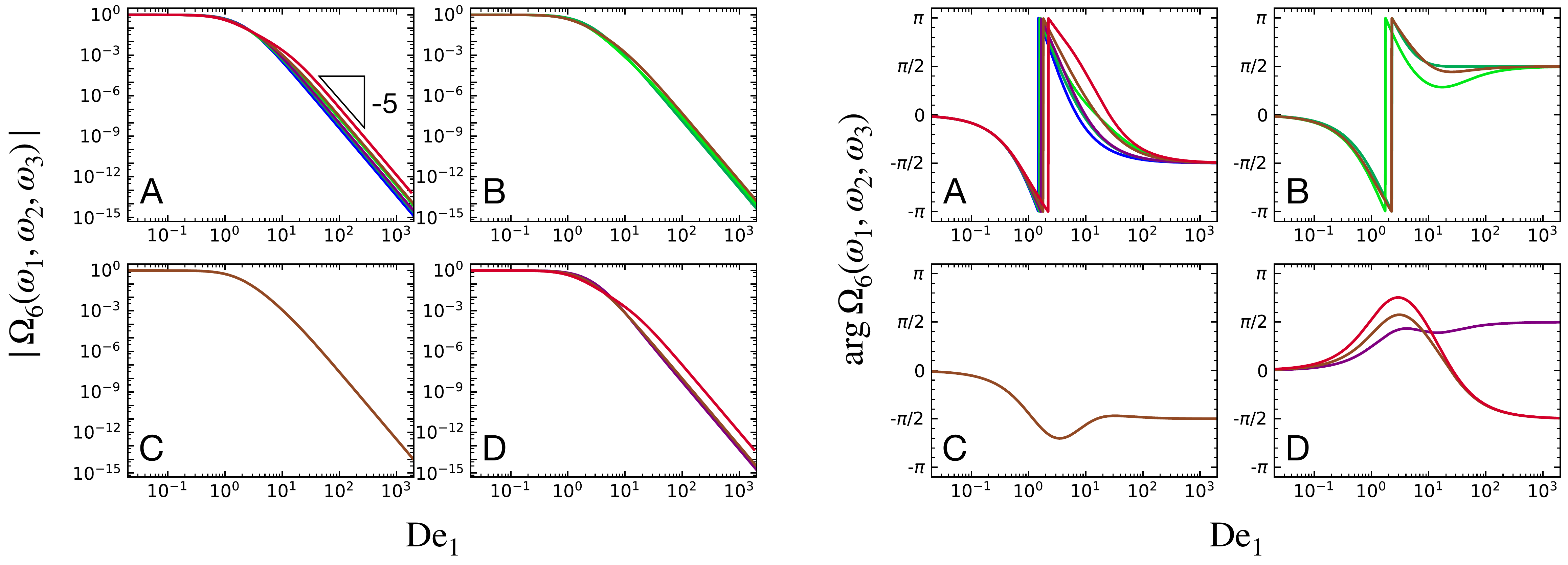}
    \caption{The magnitude (left) and phase (right) of the function $\Omega_6(\omega_1,\omega_2,\omega_3)$ with $\lambda_2 = 0$ along the manifolds studied by a three-tone frequency sweep with $\{n_1,n_2,n_3\} = \{1,4,16\}$, separated into subspaces A, B, C, and D within the MAPS domain. Colors distinguish different barycentric coordinates on the constant $L^1$-norm projection of the MAPS domain, corresponding to the locations in Figure \ref{fig:plotting}b.}
    \label{fig:bode_omega6}
\end{figure*}

Figures \ref{fig:bode_omega1} through \ref{fig:bode_omega6} employ this visualization scheme, depicting Bode plots of the magnitude and phase of the functions $\Omega_n(\omega_1,\omega_2,\omega_3)$ with $\lambda_2 = 0$ along the manifolds studied by a continuous three-tone frequency sweep with $\{n_1,n_2,n_3\} = \{1,4,16\}$. The abscissa is made dimensionless as a Deborah number with respect to the $L^1$-norm of the three-frequency coordinate $(\omega_1,\omega_2,\omega_3)$:
\begin{equation}
    \mathrm{De}_1 \equiv \lambda_1\sum_j|\omega_j|.
\end{equation}
In Figures \ref{fig:bode_omega1} through \ref{fig:bode_omega6}, only the functions $\Omega_n(\omega_1,\omega_2,\omega_3)$ for $n = 1\ldots 6$ that are active in the quadratic Maxwell model are shown, both for the sake of brevity and because many of the constitutive models presented in Table \ref{tab:sub_models} fall within this more compact framework.

Figures \ref{fig:bode_omega1} through \ref{fig:bode_omega6} illustrate that the six basis functions that comprise the third order complex viscosity for the quadratic Maxwell model are clearly distinguishable in three-tone MAPS experiments. Firstly, many of these complex functions can be distinguished by the high-frequency ($\mathrm{De}_1 \gg 1$) behavior of their magnitude. The function $\Omega_1(\omega_1,\omega_2,\omega_3)$, for instance, is the only function whose magnitude exhibits a high frequency dependence of $\mathrm{De}_1^{-2}$, and likewise $\Omega_6(\omega_1,\omega_2,\omega_3)$ is the only one whose magnitude scales like $\mathrm{De}_1^{-5}$ in this limit. Though the pair $\Omega_2(\omega_1,\omega_2,\omega_3)$ and $\Omega_3(\omega_1,\omega_2,\omega_3)$ and the pair $\Omega_4(\omega_1,\omega_2,\omega_3)$ and $\Omega_5(\omega_1,\omega_2,\omega_3)$ are not clearly distinguishable from their magnitudes alone, the dependence of the phase angle of these functions on $\mathrm{De}_1$ is quite distinct. One simple way to distinguish $\Omega_2(\omega_1,\omega_2,\omega_3)$ from $\Omega_3(\omega_1,\omega_2,\omega_3)$, for example, is to note that all of the curves in subspaces B, C, and D for $\Omega_2(\omega_1,\omega_2,\omega_3)$ asymptotically approach a phase angle of $-\pi/2$ at high frequency, while many of the curves in the same subspaces for $\Omega_3(\omega_1,\omega_2,\omega_3)$ approach a phase of $\pi/2$. Similarly, the high-frequency limits of the phase for all curves in subspaces C and D for $\Omega_4(\omega_1,\omega_2,\omega_3)$ from $\Omega_5(\omega_1,\omega_2,\omega_3)$ are out of phase by $\pi$. Because the behavior of the phase angle is similar for all basis functions at low frequencies, this distinct high-frequency behavior implies that all basis functions are linearly independent.

Because the features of each basis function $\Omega_n(\omega_1,\omega_2,\omega_3)$ that are revealed by the three-tone MAPS experiment studied here are clearly distinct, it follows that this experimental protocol produces suitable data for model identification within the framework of the quadratic Maxwell model. Moreover, the 19 complex data points that can be obtained from just a single three-tone experiment are more than sufficient to determine the six coefficients $A_n$ in the expansion of the third order complex viscosity for the quadratic Maxwell model (equation \ref{eq:simpler_solution}).

\subsection{Single-Tone Oscillatory Shear}
\label{sec:maos}

\begin{figure*}[ht!]
    \centering
    \includegraphics[width=\textwidth]{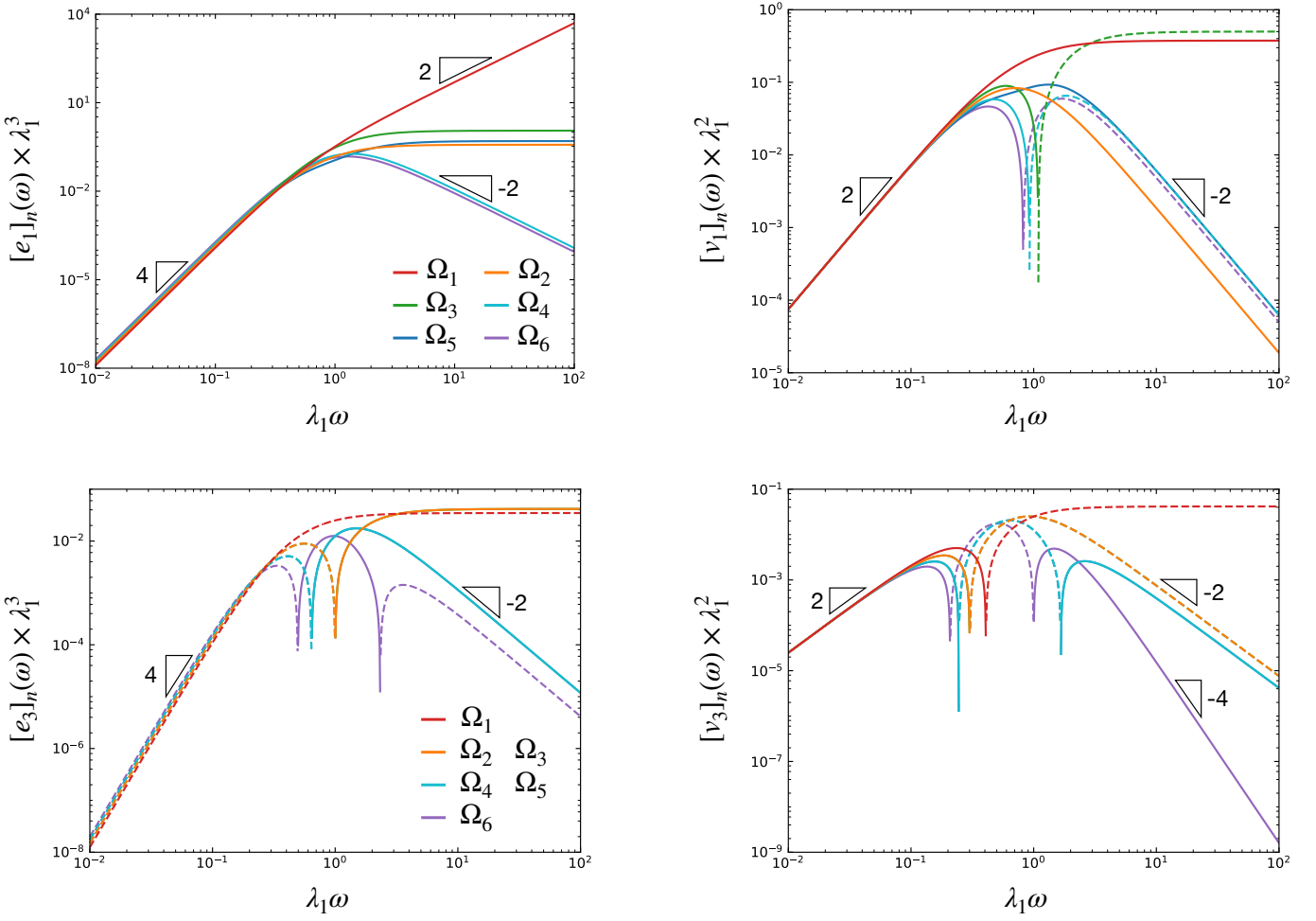}
    \caption{The contributions of each basis function $\Omega_n(\omega_1,\omega_2,\omega_3)$ to the intrinsic nonlinear functions defined for MAOS: $[e_1](\omega)$, $[v_1](\omega)$, $[e_3](\omega)$, and $[v_3](\omega)$, with $\lambda_2 = 0$. The contributions $[e_m]_n$ and $[v_m]_n$, and the frequency coordinate $\omega$, have been made dimensionless by the relaxation time $\lambda_1$. Positive values are indicated with solid lines, and negative values with dashed lines.}
    \label{fig:maos_signatures}
\end{figure*}

One of the most commonly applied deformation protocols in shear rheometry is the single-tone sinusoidal oscillation:
\begin{equation}
    \gamma(t) = \gamma_0 \sin(\omega t).
    \label{eq:single_tone}
\end{equation}
Small amplitude oscillatory shear (SAOS) experiments directly probe the linear complex viscosity, and SAOS frequency sweeps have been applied to fully characterize the linear response of complex fluids over many orders of magnitude in timescale \cite{tschoegl-1989,geri-2018}. In nonlinear rheology, a combination of frequency and amplitude sweeps in large amplitude oscillatory shear (LAOS) have been used to construct detailed ``rheological fingerprints'' for a range of materials \cite{giacomin-1993,mckinley-2008}. Recently, the weakly nonlinear version of a single-tone oscillatory experiment, called medium amplitude oscillatory shear (MAOS), has gained traction as a method for effectively studying the intrinsic nonlinearities of complex fluids \cite{ewoldt-2013}. The response of a fluid in MAOS has been defined in terms of four real-valued, intrinsic nonlinear functions:
\begin{align}
    \sigma(t) =& \gamma_0\{G'(\omega)\sin(\omega t) + G''(\omega)\cos(\omega t)\} \\
    & + \gamma_0^3\Big\{[e_1](\omega)\sin(\omega t) + \omega[v_1](\omega)\cos(\omega t) \nonumber \\
    & - [e_3](\omega)\sin(3\omega t) + \omega [v_3](\omega)\cos(3 \omega t)\Big\} + O(\gamma_0^5) \nonumber,
    \label{eq:maos}
\end{align}
where $G'(\omega)$ and $G''(\omega)$ represent the real and imaginary components of the complex modulus from linear viscoelasticity. The intrinsic nonlinear function $[e_1](\omega)$ measures the elastic component of the third order material response at the first harmonic of the input frequency, while $[v_1](\omega)$ measures the viscous component of the third order response at the first harmonic. Likewise, $[e_3](\omega)$ and $[v_3](\omega)$ measure the elastic and viscous components of the third order response at the third harmonic. These four intrinsic nonlinear functions each represent projections of the third order complex viscosity onto specific one-dimensional manifolds \cite{lennon-2020-1}:
\begin{subequations}
\label{eq:maos_all}
    \begin{align}
    [e_1](\omega) &= \frac{3\omega^3}{4} \eta^{\prime\prime}_3(\omega,-\omega,\omega),\\
    [v_1](\omega) &= \frac{3\omega^2}{4} \eta^{\prime}_3(\omega,-\omega,\omega),\\
    [e_3](\omega) &=  -\frac{\omega^3}{4} \eta^{\prime\prime}_3(\omega,\omega,\omega),\\
    [v_3](\omega) &= \frac{\omega^2}{4} \eta^{\prime}_3(\omega,\omega,\omega).
    \end{align}
\end{subequations}
Each intrinsic MAOS function relates linearly to either the real or imaginary component of the third order complex viscosity, $\eta^*_3(\omega_1,\omega_2,\omega_3) = \eta'_3(\omega_1,\omega_2,\omega_3) - i\eta''(\omega_1,\omega_2,\omega_3)$, along a particular ray in three-frequency space. Therefore, equation \ref{eq:cubic_solution} can be substituted into equation \ref{eq:maos_all} to reveal series expansions for each intrinsic nonlinear MAOS function for the cubic Maxwell model, and likewise with equation \ref{eq:simpler_solution} for the quadratic Maxwell model. With the appropriate parameter substitutions for the models defined in Table \ref{tab:sub_models}, these series expansions can be used to validate that equation \ref{eq:cubic_solution} reduces to previously known solutions for the intrinsic nonlinearities in MAOS. Indeed, we find that equations \ref{eq:cubic_solution} and \ref{eq:maos_all} along with Tables \ref{tab:sub_models} and \ref{tab:cubic_coefficients} can be used to derive the MAOS solutions for the Johnson-Segalman, Phan-Thien--Tanner, Larson, stretching and non-stretching Gaussian Rolie-Poly, and Gaussian cDCR-CS models found by Hyun and co-workers \cite{yong-2020}; the solution for the Giesekus model found by Gurnon and Wagner \cite{gurnon-2012}; and the solution for the Oldroyd 8-constant framework found by Giacomin and co-workers \cite{poungthong-2019}. That all of these solutions, which had previously been derived and represented separately, can be found as realizations of the solution for the cubic Maxwell model further demonstrates both the generality of this framework and its potential utility in model identification.

As discussed previously, the intrinsic nonlinear MAOS functions for the quadratic Maxwell model can be compactly expressed as series with contributions from each basis function $\Omega_n(\omega_1,\omega_2,\omega_3)$:
\begin{subequations}
\label{eq:maos_signatures}
\begin{align}
    [e_m](\omega) &= \eta_0 \sum_{n=1}^6 A_n [e_m]_n, \\
    [v_m](\omega) &= \eta_0 \sum_{n=1}^6 A_n [v_m]_n,
\end{align}
\end{subequations}
where each $[e_m]_n$ and $[v_m]_n$ can be obtained by replacing the terms $\eta'_3(\omega_1,\omega_2,\omega_3)$ and $\eta''_3(\omega_1,\omega_2,\omega_3)$ in equation \ref{eq:maos_all} with $\Omega'_n(\omega_1,\omega_2,\omega_3)$ and $\Omega''_n(\omega_1,\omega_2,\omega_3)$, respectively, which are the real and imaginary components of the function $\Omega_n(\omega_1,\omega_2,\omega_3) = \Omega'_n(\omega_1,\omega_2,\omega_3) - i\Omega''_n(\omega_1,\omega_2,\omega_3)$. Figure \ref{fig:maos_signatures} presents $[e_m]_n$ and $[v_m]_n$ for all six basis functions in the expansion for the quadratic Maxwell model, with $\lambda_2 = 0$.

Examination of Figure \ref{fig:maos_signatures} reveals that the basis functions $\Omega_n(\omega_1,\omega_2,\omega_3)$ do not produce distinct signatures in every intrinsic nonlinear MAOS function. In particular, at the third harmonic, the pair of functions $\Omega_2(\omega_1,\omega_2,\omega_3)$ and $\Omega_3(\omega_1,\omega_2,\omega_3)$ produce the exact same frequency-dependent elastic and viscous signatures. The pair $\Omega_4(\omega_1,\omega_2,\omega_3)$ and $\Omega_5(\omega_1,\omega_2,\omega_3)$ also produce the same signatures of the third harmonic. Thus, third harmonic data from MAOS experiments alone are insufficient to distinguish all six distinct contributions from the quadratic Maxwell model. This is a significant realization, particular because the third harmonic nonlinear material functions in MAOS are more easily measured than the first harmonic nonlinear material functions, due to the fact that nonlinearities on the first harmonic are intermixed with the linear response \cite{singh-2018}.

At the first harmonic, the elastic and viscous MAOS signatures of each function $\Omega_n(\omega_1,\omega_2,\omega_3)$ are distinct, but not necessarily easily distinguishable. In particular, the functions $\Omega_2(\omega_1,\omega_2,\omega_3)$ and $\Omega_5(\omega_1,\omega_2,\omega_3)$ produce signatures at the first harmonic that are qualitatively quite similar, as do the functions $\Omega_4(\omega_1,\omega_2,\omega_3)$ and $\Omega_6(\omega_1,\omega_2,\omega_3)$. Segregating these functions from first harmonic MAOS data alone is thus a poorly conditioned problem. Therefore, in order to distinguish all six basis functions from MAOS data, it is necessary to consider both first and third harmonic MAOS data. Because identifying the third order nonlinearities at the first harmonic in MAOS requires at least two experiments at different strain amplitudes \cite{singh-2018}, this expands the amount of experimentation necessary to build a data set for model identification purposes. Moreover, MAOS experiments at a single frequency produce only four real-valued data points, which alone are insufficient to infer the six $A_n$ coefficients in equation \ref{eq:maos_signatures}. Therefore MAOS experiments at a minimum of two frequencies are required for model identification within the framework of the quadratic Maxwell model, and even more are required for accurate model identification in more general frameworks such as the cubic Maxwell model. This is in stark contrast to the three-tone MAPS protocol, for which a single experiment provides enough distinct data for well-conditioned model identification in either of these frameworks.

\subsection{Startup of Steady Shear}
\label{sec:startup}

Oscillatory deformation protocols such as the single-tone and three-tone protocols measure material properties in the frequency-domain. Many deformation protocols exist that measure time-domain material functions instead, such as the startup of steady shear flow. The shear-rate deformation protocol in this experiment is given by a step function:
\begin{equation}
    \dot\gamma(t) = \dot\gamma_0 H(t),
    \label{eq:step_strain_rate}
\end{equation}
where $H(t)$ represents the Heaviside step function. The shear stress response to this deformation is often represented in terms of a ``transient'' or ``startup'' viscosity $\eta^+(t,\dot\gamma_0)$ \cite{bird-1987}:
\begin{equation}
    \sigma(t) = \eta^+(t,\dot\gamma_0)\dot\gamma_0.
\end{equation}
In a weakly nonlinear startup of steady shear experiment, the shear stress response can be written as an expansion in $\dot\gamma_0$:
\begin{equation}
    \sigma(t) = \eta^+_1(t)\dot\gamma_0 + \eta^+_3(t)\dot\gamma_0^3 + O(\dot\gamma_0^5),
    \label{eq:startup_viscosity}
\end{equation}
where $\eta^+_1(t)$ and $\eta^+_3(t)$ are the linear and weakly nonlinear contributions to the startup viscosity, both of which are independent of the deformation strength.

In relating $\eta^+_1(t)$ and $\eta^+_3(t)$ to the third order complex viscosity, it is convenient to define the time-domain Volterra series \cite{rivlin-1957}:
\begin{equation}
    \sigma(t) = \sum_{n=1}^{\infty}\int\ldots\int_{-\infty}^{t}G_n(t-t_1,\ldots,t-t_n)\prod_{m=1}^n \dot\gamma(t_m) dt_m,
    \label{eq:time_volterra}
\end{equation}
with kernels $G_n(t_1,\ldots,t_n)$ called the \emph{nth order relaxation moduli}, which are related to the $n$th order complex viscosities by the $n$-dimensional inverse Fourier transform:
\begin{equation}
    G_n(t_1,\ldots,t_n) = \frac{1}{(2\pi)^n}\int\ldots\int \eta^*_n(\omega_1,\ldots,\omega_n)\prod_{m=1}^n e^{i\omega_m t_m} d\omega_m.
\end{equation}
Substituting equation \ref{eq:step_strain_rate} into equation \ref{eq:time_volterra} and comparing to equation \ref{eq:startup_viscosity}, we find that:
\begin{equation}
    \eta^+_n(t) = \int\ldots\int_0^t G_n(t-t_1,\ldots,t-t_n) dt_1 \ldots dt_n.
\end{equation}

\begin{figure}
    \centering
    \includegraphics[width=\columnwidth]{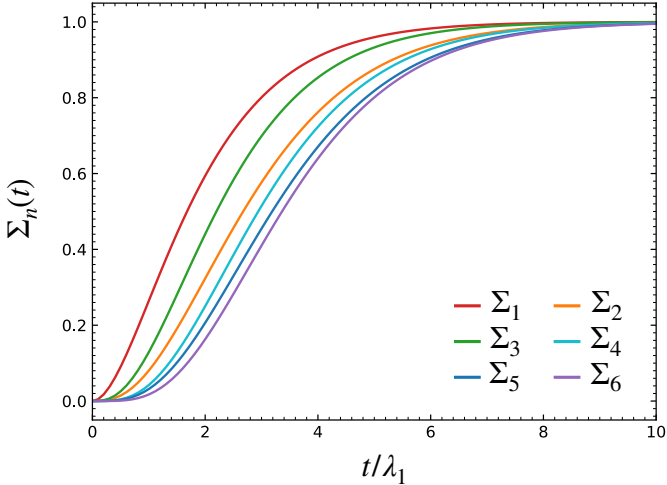}
    \caption{The basis functions $\Sigma_n(t)$ for a weakly nonlinear startup of steady shear experiment, with $\lambda_2 = 0$. The basis functions $\Sigma_n(t)$ are dimensionless quantities, while time has been made dimensionless by the relaxation time $\lambda_1$.}
    \label{fig:startup_basis}
\end{figure}

Taking the inverse Fourier transform of $\eta^*_1(\omega)$ for the generalized nonlinear Maxwell model (equation \ref{eq:eta1}) and integrating is straightforward, giving:
\begin{equation}
    \eta^+_1(t) = \eta_0\left[1 - \left(1 - \frac{\lambda_2}{\lambda_1}\right)e^{-t/\lambda_1}\right].
    \label{eq:startup_linear}
\end{equation}
Due to the linearity of the inverse Fourier transform and the integral operator, each basis function $\Omega_n(\omega_1,\omega_2,\omega_3)$ from the quadratic Maxwell model produces an independent and additive contribution to $\eta^+_3(t)$, which therefore may be expressed as a basis expansion:
\begin{equation}
    \eta^+_3(t) = \eta_0 \sum_{n=1}^6 A_n \Sigma_n(t),
    \label{eq:startup_expansion}
\end{equation}
where $\Sigma_n(t)$ is the contribution arising from the basis function $\Omega_n(\omega_1,\omega_2,\omega_3)$. The functions $\Sigma_n(t)$ are presented in the Supporting Information, and each $\Sigma_n(t)$ is shown in Figure \ref{fig:startup_basis}. Note that equations \ref{eq:startup_linear} and \ref{eq:startup_expansion}, along with the expressions for each $\Sigma_n(t)$ provided in the Supporting Information, reduce to the proper weakly nonlinear expansions of previously derived solutions for $\eta^+(t,\dot\gamma)$ for the corotational Maxwell and Jeffreys models \cite{bird-1987}. It also reduces to the cubic expansion of the solution recently derived for the Oldroyd 8-constant framework when $\lambda_2 = 0$ \cite{saengow-2019}, but for a nonzero value of $\lambda_2$ reveals an additional contribution from the impulse at $t = 0$. To validate this expansion, we provide a full derivation of $\eta^+(t,\dot\gamma)$ for the Oldroyd 8-constant framework in the Supporting Information that includes this impulse feature.

Figure \ref{fig:startup_basis}, firstly, demonstrates the stark difference in the dimensionality of data coming from a weakly nonlinear startup of steady shear experiment as compared to MAOS or three-tone MAPS experiments. In particular, a weakly nonlinear startup experiment provides only one-dimensional (real-valued) data on the real line defined by time. MAOS, on the other hand, accesses four-dimensional real-valued data on the real line defined by frequency. A three-tone MAPS experiment results in 19-dimensional complex-valued data, equivalent to 38-dimensional real-valued data, along the frequency line. 

Though it is possible to obtain data across the real-line in a startup experiment with just two experiments, as opposed to the multiple experiments comprising a frequency sweep in oscillatory measurements, the severely reduced dimensionality of the output clearly limits the usefulness of weakly nonlinear startup data in model identification. Though the six basis functions $\Sigma_n(t)$ are distinct in Figure \ref{fig:startup_basis}, they are qualitatively very similar, and in certain cases almost quantitatively identical. Inversion of equation \ref{eq:startup_expansion} is therefore very poorly conditioned, and accurately distinguishing the six coefficients $A_n$ from even slightly noisy data may be practically impossible.

\section{Model Identification with Weakly Nonlinear Data}
\label{sec:model_identification}

To study which of the three deformation protocols outlined in the previous section is most suitable for the model identification task outlined in section \ref{sec:maps_response}, we will restrict ourselves to model identification within the quadratic Maxwell model, which is based on the linear system defined by equation \ref{eq:simpler_solution} with $N = 6$. In this context, the model identification problem for each of the three deformation protocols can be expressed in the form:
\begin{equation}
    \textbf{y} = \textbf{M}\textbf{A},
    \label{eq:linear_problem}
\end{equation}
where $\textbf{y}$ is an $N \times 1$ data vector, $\textbf{A}$ is a $6 \times 1$ vector whose elements are related to the coefficients $A_n$ defined by equation \ref{eq:A_coeff} and Table \ref{tab:cubic_coefficients}, and $\textbf{M}$ is a $N \times 6$ matrix representing a suitable basis for the data, whose elements are related to the basis functions $\Omega_n(\omega_1,\omega_2,\omega_3)$. We will assume that, in formulating this problem, the linear viscoelastic response of the material is known and has been fit to a single-mode Maxwell model with parameters $\eta_0$ and $\lambda_1$. Thus, the vectors $\textbf{y}$ and $\textbf{A}$, and the matrix $\textbf{M}$, can be made dimensionless such that the elements of $\textbf{A}$ correspond to the scaled lumped coefficients of the quadratic Maxwell model:
\begin{equation}
    \textbf{A} = \frac{1}{\lambda_1^2}\left[\begin{array}{cccccc}
        A_1 & A_2 & A_3 & A_4 & A_5 & A_6
    \end{array}\right]^T.
\end{equation}
In the following three subsections, we relate $\textbf{y}$ and $\textbf{M}$ to these basis functions and to the data obtained by each of the three deformation protocols explored previously: MAPS, MAOS, and shear start-up. Though we focus here on the model identification problem with a single mode described by the quadratic Maxwell model, the same analysis may be directly applied to a fit to $K$ such modes by concatenating together one matrix $\textbf{M}_k$ per mode to form $\textbf{M}$ and one vector $\textbf{A}_k$ per mode to form $\textbf{A}$, and defining suitable parameters $\eta_0$ and $\lambda_1$ for non-dimensionalization.

\subsection{Model Identification with Three-Tone MAPS Data}

A single three-tone MAPS experiment, as discussed in Section \ref{sec:three_tones} and in more detail by Lennon et. al. \cite{lennon-2020-2}, produces discrete measurements of the third order MAPS response functions at 19 distinct points in three-frequency space: $(\omega_1,\omega_2,\omega_3)$. Because of the generality of MAPS rheology, the measured data can be readily converted to any of the four defined MAPS response functions: the third order complex viscosity $\eta^*_3(\omega_1,\omega_2,\omega_3)$, the third order complex modulus $G^*_3(\omega_1,\omega_2,\omega_3)$, the third order complex compliance $J^*_3(\omega_1,\omega_2,\omega_3)$, and the third order complex fluidity $\phi^*_3(\omega_1,\omega_2,\omega_3)$. In strain-controlled rheometry, it is more natural to express the data in terms of one of the former two response functions. For instance, the third order complex viscosity can be measured in order to construct a regression problem that matches exactly the form of equation \ref{eq:simpler_solution}. In this case, the elements of $\textbf{y}$ in equation \ref{eq:linear_problem} are either the real or imaginary component of the measured third order complex viscosity, and elements of $\textbf{M}$ are each values of a basis function $\Omega_n(\omega_1,\omega_2,\omega_3)$ at distinct points in three-frequency space. While this is a well-posed problem, there is a practical concern associated with this formulation. In particular, we have seen that high-frequency ($\mathrm{De}_1 > 1$) data is required to distinguish between the basis functions $\Omega_n(\omega_1,\omega_2,\omega_3)$. At high frequencies, however, the magnitude of these basis functions decays rapidly with the frequency $L^1$-norm, thus individual values in the matrix $\textbf{M}$ will be of vastly different scales. The resulting inverse problem of determining the least-squares fitting coefficients $\textbf{A}$ is therefore likely to be poorly conditioned, and the regressed values of $\textbf{A}$ highly sensitive to even small uncertainty in the third order complex viscosity data.

The condition number of the matrix $\textbf{M}$ can be improved, however, by instead considering the third order complex modulus, $G^*_3(\omega_1,\omega_2,\omega_3)$. The third order complex modulus in MAPS rheology can be computed from the third order complex viscosity:
\begin{equation}
    G^*_3(\omega_1,\omega_2,\omega_3) = \left(\prod_{m=1}^3 i\omega_m\right)\eta^*_3(\omega_1,\omega_2,\omega_3).   
\end{equation}
Thus, in the quadratic Maxwell model, $G^*_3(\omega_1,\omega_2,\omega_3)$ is related to the coefficients $A_n$ by:
\begin{equation}
    G^*_3(\omega_1,\omega_2,\omega_3) = -i\omega_1 \omega_2 \omega_3 \sum_{n=1}^6 \Omega_n(\omega_1,\omega_2,\omega_3)A_n.
    \label{eq:complex_modulus}
\end{equation}
In this case, we may express the data vector $\textbf{y}$ as:
\begin{subequations}
\begin{equation}
    \textbf{y} = \left[\begin{array}{c}
        \textbf{y}_{Re} \\
        \textbf{y}_{Im} 
    \end{array}\right],
\end{equation}
\begin{equation}
    \left\{\begin{array}{l}
        (y_{Re})_q = G'_3(\omega_{1,q}, \omega_{2,q}, \omega_{3,q}) \times (\lambda_1/\eta_0) \\
        (y_{Im})_q = G''_3(\omega_{1,q}, \omega_{2,q}, \omega_{3,q}) \times (\lambda_1/\eta_0)
    \end{array}\right.
\end{equation}
\end{subequations}
where $q = 1, \ldots, Q$ represent measurements of $G^*_3(\omega_1,\omega_2,\omega_3) = G'_3(\omega_1,\omega_2,\omega_3)$ $+ iG''_3(\omega_1,\omega_2,\omega_3)$ at $Q$ distinct three-frequency coordinates. Because a three-tone MAPS experiment at a single fundamental frequency produces 19 distinct measurements of the MAPS response functions, three-tone MAPS experiments at $p$ different fundamental frequencies will result in $Q = 19p$ measurements. Therefore, $\textbf{y}$ is a $38p$-dimensional vector; thus, even a single ($p = 1$) three-tone experiment produces more than enough data to infer each of the six coefficients $A_n$ in the quadratic Maxwell model.

Based on equation \ref{eq:complex_modulus}, the matrix $\textbf{M}$ has the construction:
\begin{subequations}
\label{eq:M_MAPS}
\begin{equation}
    \textbf{M} = \left[\begin{array}{c}
        \textbf{M}_{Re} \\
        \textbf{M}_{Im} 
    \end{array}\right],
\end{equation} 
\begin{equation}
    \left\{\begin{array}{l}
        (M_{Re})_{q,n} = -\lambda_1^3 \omega_{1,q} \omega_{2,q} \omega_{3,q} \Omega''_n(\omega_{1,q}, \omega_{2,q}, \omega_{3,q}) \\
        (M_{Im})_{q,n} = -\lambda_1^3 \omega_{1,q} \omega_{2,q} \omega_{3,q} \Omega'_n(\omega_{1,q}, \omega_{2,q}, \omega_{3,q})
    \end{array}\right.
\end{equation}
\end{subequations}
for $n = 1, \ldots, 6$. Because elements of $\textbf{M}$ are now multiplied by each of the frequency coordinates, these elements do not decay as strongly with the frequency $L^1$-norm, which results in a better conditioned matrix $\textbf{M}$. We will shortly explore how the details of the three-tone MAPS experiments used to formulate this matrix affect its condition number.

In addition to the formulation of the problem in terms of the third order complex modulus being more well-conditioned than the formulation in terms of the third order complex viscosity, it is also perhaps the more natural representation for strain-controlled rheometry. Commercial rheometers commonly measure the shear stress and shear strain directly, rather than the shear strain rate. The shear stress and strain are related directly by the third order complex modulus as shown previously. Thus, measurement errors should correlate more directly to values of the third order complex modulus than to the third order complex viscosity, and propagation of uncertainty to the coefficients of the quadratic Maxwell model is most naturally examined by using the $G^*_3(\omega_1,\omega_2,\omega_3)$ in the regression problem.

Though we have chosen a strain-controlled representation of the regression problem here, it is possible to instead choose to represent the problem using the stress-controlled MAPS response functions, such as the third order complex compliance. Lennon et. al. \cite{lennon-2020-2} have demonstrated the advantages of stress-controlled rheometry in multi-tone MAPS measurements, which directly measure $J^*_3(\omega_1,\omega_2,\omega_3)$. Because this function is, loosely speaking, the inverse of $G^*_3(\omega_1,\omega_2,\omega_3)$, formulating the regression problem in terms of $J^*_3(\omega_1,\omega_2,\omega_3)$ results in a similarly-conditioned problem to the previous formulation in terms of $G^*_3(\omega_1,\omega_2,\omega_3)$. For the remainder of this work, however, we will continue our discussion in the context of strain control, and simply note that the analysis may be extended straightforwardly to the stress-controlled case.

\subsection{Model Identification with MAOS Data}

Strain-controlled MAOS experiments directly measure the four intrinsically nonlinear MAOS functions: $[e_1](\omega)$, $[v_1](\omega)$, $[e_3](\omega)$ and $[v_3](\omega)$. Measurement of the third harmonic functions $[e_3](\omega)$ and $[v_3](\omega)$ requires, at minimum, one experiment at a single amplitude, while measurement of the first harmonic functions $[e_1](\omega)$ and $[v_1](\omega)$ requires experiments at a minimum of two amplitudes. We have seen in Section \ref{sec:maos} that certain basis functions are degenerate on the third harmonic in MAOS, therefore it is necessary to measure the first harmonic material functions in order for the regression problem to be non-singular. The data vector $\textbf{y}$ in MAOS is therefore:
\begin{subequations}
\begin{equation}
    \textbf{y} = \left[\begin{array}{c}
        \textbf{y}_{[e_1]} \\
        \textbf{y}_{[v_1]} \\
        \textbf{y}_{[e_3]} \\
        \textbf{y}_{[v_3]}
    \end{array}\right],
\end{equation}
\begin{equation}
    \left\{\begin{array}{l}
        (y_{[e_1]})_q = [e_1](\omega_q) \times (\lambda_1/\eta_0)\\
        (y_{[v_1]})_q = \omega_q [v_1](\omega_q) \times (\lambda_1/\eta_0)\\
        (y_{[e_3]})_q = [e_3](\omega_q) \times (\lambda_1/\eta_0)\\
        (y_{[v_3]})_q = \omega_q [v_3](\omega_q) \times (\lambda_1/\eta_0)
    \end{array}\right.
\end{equation}
\end{subequations}
with $q = 1, \ldots, Q$. A single MAOS experiment measures these functions at only a single frequency, thus $p$ MAOS experiments will result in $Q = p$ measurements of each. Thus, $\textbf{y}$ is a $4p$-dimensional vector, and we see that MAOS experiments at a minimum of $p = 2$ frequencies (with a minimum of two input amplitudes per frequency) are required to uniquely determine all six coefficients $A_n$.

From equations \ref{eq:simpler_solution} and \ref{eq:maos_all} we can construct the matrix $\textbf{M}$:
\begin{subequations}
\label{eq:M_MAOS}
\begin{equation}
    \textbf{M} = \left[\begin{array}{c}
        \textbf{M}_{[e_1]} \\
        \textbf{M}_{[v_1]} \\
        \textbf{M}_{[e_3]} \\
        \textbf{M}_{[v_3]}
    \end{array}\right],
\end{equation} 
\begin{equation}
    \left\{\begin{array}{l}
        (M_{[e_1]})_{q,n} = (3/4) (\lambda_1 \omega_q)^3 \Omega''_n(\omega_{q}, -\omega_{q}, \omega_{q}) \\
        (M_{[v_1]})_{q,n} = (3/4) (\lambda_1 \omega_q)^3 \Omega'_n(\omega_{q}, -\omega_{q}, \omega_{q}) \\
        (M_{[e_3]})_{q,n} = -(1/4) (\lambda_1 \omega_q)^3 \Omega''_n(\omega_{q}, \omega_{q}, \omega_{q}) \\
        (M_{[v_3]})_{q,n} = (1/4) (\lambda_1 \omega_q)^3 \Omega'_n(\omega_{q}, \omega_{q}, \omega_{q}) \\
    \end{array}\right..
\end{equation}
\end{subequations}

\subsection{Model Identification with Shear Startup Data}

The case of the startup of steady shear differs from the three-tone MAPS experiments and MAOS experiments, in that it is possible to measure the governing weakly nonlinear material function $\eta^+_3(t)$, continuously across its domain with only a few experiments. In particular, isolating this weakly nonlinear component of the startup viscosity requires at minimum two startup experiments of the same duration and sampling frequency, but different shear rates $\dot\gamma_0$. With the values of $\eta^+_3(t)$ determined at $Q$ discrete points $t_q$, the data vector $\textbf{y}$ is simply:
\begin{equation}
    (y)_q = \eta^+_3(t_q)/\eta_0,
\end{equation}
for $q = 1, \ldots, Q$, and the matrix $\textbf{M}$ is found directly from equation \ref{eq:startup_expansion}:
\begin{equation}
    (M)_{q,n} = \lambda_1^2 \Sigma_n(t_q),
    \label{eq:M_startup}
\end{equation}
where $\Sigma_n$ are the basis functions for the quadratic Maxwell model in shear startup, defined in Section \ref{sec:startup}. Because $Q$ can be controlled by either the duration of the experiment or the sampling rate, it is possible to obtain enough data for the regression problem by measuring $\eta^+_3(t)$ only once, requiring just two experiments at different strain rates $\dot\gamma_0$. However, because the basis functions in shear startup behave quite similarly across the entire time domain, this regression problem is naturally ill-posed. Even the combination of long experiments and high sampling rates, resulting in very high values of $Q$, cannot decrease the condition number of the matrix $\textbf{M}$ substantially.

\subsection{Propagation of Uncertainty}

Model identification can in principle be performed using any of the previously considered data sets, and in the case where there is no uncertainty in the data, each of these data sets will produce the same result for a given material. In reality, however, uncertainty in the data will exist, and this uncertainty will not propagate to the results identically for data obtained by different experimental protocols. Therefore, our interests now turn to evaluating and comparing how uncertainty in experimental data will propagate to uncertainty in the regressed values of the coefficients $A_n$ for each of the experimental protocols considered previously in this section.

An important metric for how uncertainty propagates in a linear problem of the form $\textbf{y} = \textbf{M}\textbf{A}$ is the condition number of the matrix $\textbf{M}$. The condition number of $\textbf{M}$ represents an upper bound for the ratio of the magnitude of relative errors in the regressed vector $\textbf{A}$ and the relative error in the data vector $\textbf{y}$, where the magnitude is defined with respect to a suitable matrix norm. The condition number with respect to the $L^2$-norm, for instance, can be expressed in terms of the largest and smallest singular values of the matrix $\textbf{M}$:
\begin{equation}
    \mathrm{cond}_2(\textbf{M}) = \frac{s_1}{s_6},
    \label{eq:condition}
\end{equation}
where $s_i$, $i = 1,\ldots,6$, is the $i$th largest singular value of $\textbf{M}$.

To study the effect of the experimental protocol on the minimum condition number achievable, we will consider a few specific experimental designs. Firstly, we will consider two different three-tone MAPS frequency sweeps, with input tone sets $\{n_1,n_2,n_3\} = \{5,6,9\}$ and $\{1,4,16\}$. For these three-tone MAPS frequency sweeps, and for the MAOS frequency sweeps, we will allow the dimensionless fundamental frequency $\mathrm{De} \equiv \lambda_1\omega$ to take on $p$ distinct values spaced logarithmically between a value $\mathrm{De}_0 \equiv \lambda_1\omega_0$ and a value $\mathrm{De}_0 10^b$. For shear startup experiments, we will take data at $Q$ discrete time points spaced linearly between $t = 0$ and $t = t_f/\lambda_1$. For the three-tone MAPS and MAOS sweeps with specific values of $p$, $\mathrm{De}_0$, and $b$, the matrix $\textbf{M}$ can be constructed via equations \ref{eq:M_MAPS} and \ref{eq:M_MAOS}, respectively, and for specific values of $Q$ and $t_f/\lambda_1$ in shear startup, $\textbf{M}$ can be constructed from equation \ref{eq:M_startup}.

Firstly, we find that even for very high values of $t_f$ and $Q$, the condition number of $\textbf{M}$ for a shear startup experiment remains quite high, at a value near $10^{16}$. Thus, even small relative uncertainty in shear startup data may result in extremely large relative uncertainty in the regressed coefficients $A_n$. This is unsurprising, since the functions $\Sigma_n(t)$ in Figure \ref{fig:startup_basis} are quite similar, making the columns of $\textbf{M}$ nearly identical such that $s_6 \approx 0$. Weakly nonlinear shear startup, therefore, is very poorly conditioned for model identification in this context, and therefore we will not consider it any further.

\begin{figure*}
    \centering
    \includegraphics[width=\textwidth]{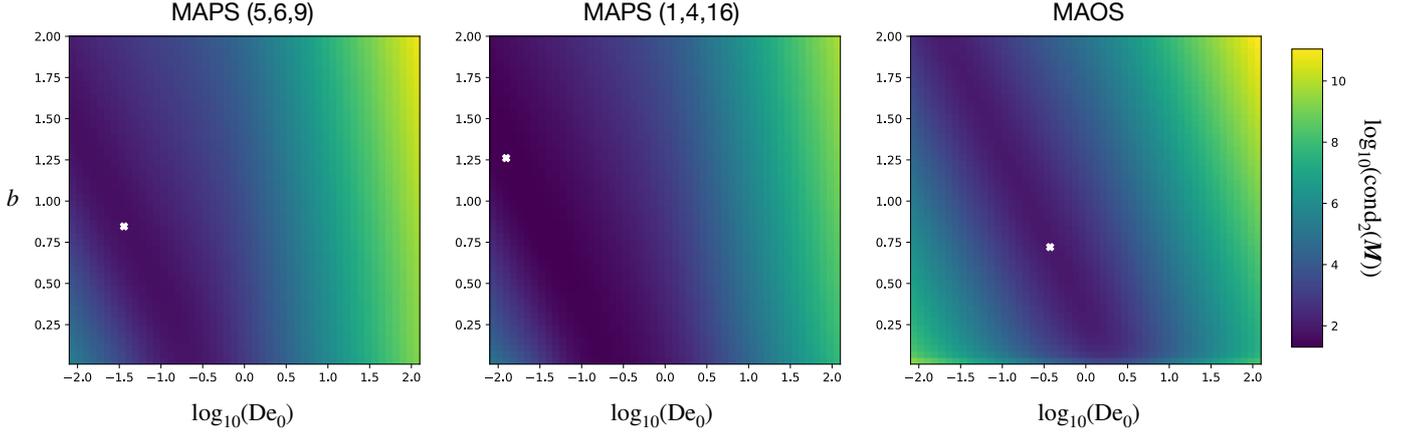}
    \caption{The $L^2$ condition number of the matrix $\mathrm{M}$ for the three-tone MAPS frequency sweep experimental protocol with $\{n_1,n_2,n_3\} = \{5,6,9\}$ and $\{1,4,16\}$, and for the MAOS frequency sweep protocol, as a function of the initial frequency $\mathrm{De}_0 \equiv \lambda_1\omega_0$ and the frequency sweep bandwidth $b = \log_{10}(\omega_{max}/\omega_0)$. The local minimum for each protocol is marked with a white cross.}
    \label{fig:cond_minimum}
\end{figure*}

We next consider the optimal experimental design for three-tone MAPS and MAOS experiments. Though we could in principle optimize the condition number for these protocols over values of $p$, $\mathrm{De}_0$, and $b$ simultaneously, we would find that no local minima exist, because the condition number decreases monotonically with $p$. Moreover, large values of $p$ would result in excessively large experimentation time. Therefore, we instead consider minimization of the condition number at a fixed value of $p = 10$, representing an extensive, but not excessive, frequency sweep. Figure \ref{fig:cond_minimum} shows the variation in the condition number of $\textbf{M}$ over a range of $\mathrm{De}_0$ and $b$ for each of the two three-tone MAPS frequency sweeps and for the MAOS frequency sweep. The local minimum, computed using Nelder-Mead optimization, is shown on each plot with a white cross. For the $\{5,6,9\}$ three-tone MAPS experiment, the minimum condition number of 49.7 occurs at $(\mathrm{De}_0, b) = (10^{-1.44},0.846)$; for the $\{1,4,16\}$ three-tone MAPS experiment, the minimum condition number of 20.0 occurs at $(\mathrm{De}_0, b) = (10^{-1.90},1.26)$; and for the MAOS experiment, the minimum of 91.5 occurs at $(\mathrm{De}_0, b) = (10^{-0.430},0.721)$.

From Figure \ref{fig:cond_minimum}, it is clear that the three-tone MAPS frequency sweeps are better conditioned for the model identification problem than the MAOS frequency sweep. In particular, the three-tone MAPS sweep with input tones $\{1,4,16\}$ produces a minimal condition number that is more than a factor of four smaller than the minimal condition number for the MAOS experiment. Moreover, this three-tone experiment results in a much wider basin over which the condition number is small compared to the MAOS experiment, thus providing additional flexibility in the experimental design. Though the three-tone MAPS sweep with input tones $\{5,6,9\}$ is not quite as well-conditioned as the sweep with input tones $\{1,4,16\}$, it still produces a minimal condition number that is a factor of two smaller than the minimal condition number for MAOS. These results are evidence that the dimensionality and diversity of the data coming from three-tone MAPS experiments, and in particular from the experiment with widely spaced input tones, is especially well suited for the problem of model identification.

\begin{figure*}[ht!]
    \centering
    \includegraphics[width = \textwidth]{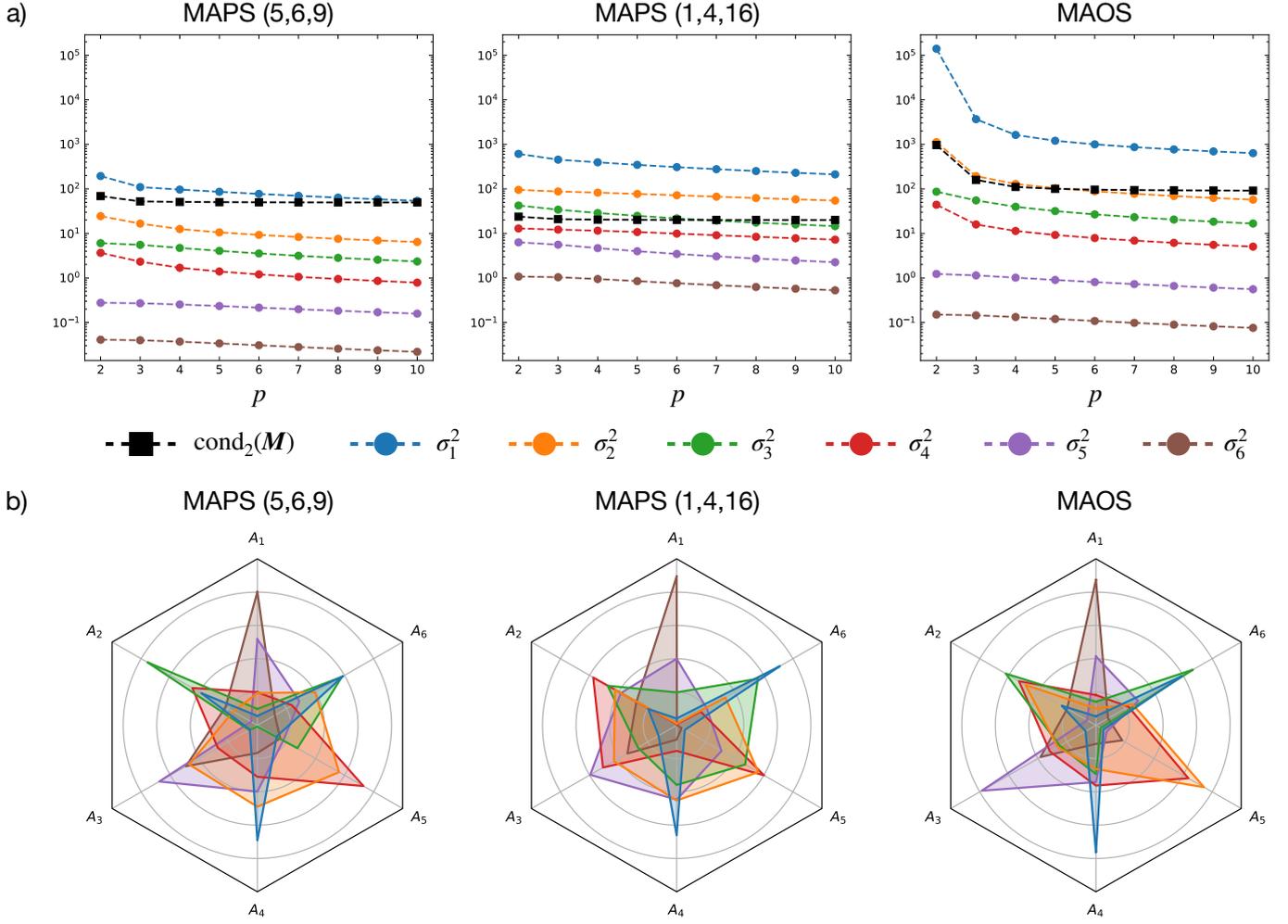}
    \caption{The eigendecomposition of the covariance matrix, computed at the optimal values of the bandwidth ($b$) and smallest frequency ($\mathrm{De}_0$) of MAPS and MAOS frequency sweeps computed in Figure \ref{fig:cond_minimum}. (a) The square of the eigenvalues of \textbf{K} (colored circles) and the condition number of \textbf{K} (black squares) as a function of the number of points $p$ in the frequency sweeps. (b) Radar plots of the components of the corresponding eigenvectors (by color) to the eigenvalues shown in (a) with $p = 10$.}
    \label{fig:eig_spectrum}
\end{figure*}

To further investigate the extent to which each of the previous experimental protocols propagates uncertainty into the regressed value of $A_n$, we may vary the number of frequencies in the frequency sweep $p$ at constant $\mathrm{De}_0$ and $b$. In Figure \ref{fig:eig_spectrum}a, we vary $p$ from 2 to 10, fixing the values of $\mathrm{De}_0$ and $b$ at the optimal coordinates found in Figure \ref{fig:cond_minimum}. This ensures that we are comparing the experimental protocols at the local optimum for each, rather than at points which may be sub-optimal for particular protocols. Figure \ref{fig:eig_spectrum}a reveals that, for both of the three-tone MAPS experiments, the condition number decreases very slightly from $p = 2$ to $p = 3$, but remains otherwise quite flat. For the MAOS experiments, on the other hand, the condition number does not reach a plateau until $p = 4$ or 5, nearly twice the number of fundamental frequencies that need be considered to reach the optimal plateau as for three-tone MAPS experiments. Moreover, we have previously noted that obtaining the necessary first harmonic data in MAOS requires at least two experiments with different amplitude at the same frequency, while the full 38-dimensional MAPS data set can be obtained in principle from a single experiment. Not only do the three-tone MAPS experiments provide a better-conditioned model identification problem than MAOS experiments, they do so with substantially less experimentation.

Though the condition number provides a useful metric for the potential severity of uncertainty magnification in the linear least-squares problem, it is instructive to examine the mechanics of uncertainty propagation in more detail. We may do so by studying the covariance matrix for the least-squares problem:
\begin{equation}
    \textbf{K} = \sigma_y^2(\textbf{M}^{T}\textbf{M})^{-1}.
\end{equation}
Assuming that the uncertainty in each input data point is normally distributed with variance $\sigma_y^2$, the elements of this covariance matrix represent the variances and covariances of a 6-dimensional normal distribution for the values of the coefficients $A_n$. We may diagonalize this covariance matrix:
\begin{subequations}
\begin{equation}
    \textbf{K} = \textbf{V}\textbf{D}\textbf{V}^T,
\end{equation}
\begin{equation}
    \textbf{V} = [\begin{array}{cccccc}
    \textbf{v}_1 & \textbf{v}_2 & \textbf{v}_3 & \textbf{v}_4 & \textbf{v}_5 & \textbf{v}_6
    \end{array}],
\end{equation}
\begin{equation}
    \mathrm{diag}(\textbf{D}) = [\begin{array}{cccccc}
    \sigma_1^2 & \sigma_2^2 & \sigma_3^2 & \sigma_4^2 & \sigma_5^2 & \sigma_6^2
    \end{array}],
\end{equation}
\end{subequations}
with the eigenvectors $\textbf{v}_i$ representing an alternative basis with uncorrelated uncertainties, and the eigenvalues $\sigma_i^2$ representing the magnitude of uncertainty in the direction of $\textbf{v}_i$. If we choose $\sigma_i^2$ to represent the $i$th largest eigenvalue, then these eigenvalues are related directly to the singular values of the matrix $\textbf{M}$:
\begin{equation}
    \sigma_i^2 = s_{N+1-i}^{-2}.
\end{equation}

This spectral decomposition of the covariance matrix firstly provides a simple geometric interpretation to the definition of the $L^2$-norm condition number in equation \ref{eq:condition}, which may be rewritten as:
\begin{equation}
    \mathrm{cond}_2(\textbf{M}) = \frac{\sigma_1}{\sigma_6}.
\end{equation}
In the context of the covariance matrix, $\sigma_1$ represents the maximum factor by which the $L^2$-norm of a vector may be stretched by $\textbf{K}$, while $\sigma_6$ represents the minimum factor by which the $L^2$-norm of a vector may be stretched by $\textbf{K}$. Therefore, the ratio of these eigenvalues indeed represents the worst case scenario in which the data vector $\textbf{y}$ is minimally stretched while the uncertainty vector $\Delta\textbf{y}$ is maximally stretched. Secondly, the spectral decomposition of the covariance matrix reveals the principal directions in which the uncertainty is concentrated. Because we have seen that some of the basis functions $\Omega_n(\omega_1,\omega_2,\omega_3)$ behave similarly, particularly near the MAOS vertices, it is likely that uncertainties in the coefficients $A_n$ will be more highly correlated. 

In Figure \ref{fig:eig_spectrum}a, we present the values of all eigenvalues of the covariance matrix as a function of $p$ at the optimal values of $(\mathrm{De}_0,b)$ for each experimental protocol. This shows that, in general, the eigenvalues can be decreased without bound by increasing the amount of data collected. We also see from this figure that the eigenvalues are lowest for the three-tone MAPS experiment with input tones $\{5,6,9\}$, and largest for the MAOS experiment. This indicates that, if the uncertainty in the data $\sigma_y^2$ is constant across experiments, rather than scaled relative to the magnitude of the data, the $\{5,6,9\}$ three-tone experiment actually will result in the smallest uncertainty in the regressed coefficients $A_n$. In this case, both of the three-tone MAPS experiments still result in lower uncertainties than comparable MAOS experiments. Lastly, this figure demonstrates that the largest eigenvalue is, in most cases, an order of magnitude or more larger than the second largest eigenvalue. Therefore, the uncertainty is especially concentrated in the direction of the first eigenvector, $\textbf{v}_1$.

In Figure \ref{fig:eig_spectrum}b, we depict the direction of each eigenvector $\textbf{v}_i$ of the covariance matrix for $p = 10$ and the values of $(\mathrm{De}_0,b)$ from Figure \ref{fig:eig_spectrum}a using radar plots. In these plots, the magnitude of the $n$th component of $\textbf{v}_i$ is shown by the vertex of the irregular hexagon, whose color corresponds to the $i$th eigenvalue, along the axis in the direction labelled $A_n$. These radar plots clearly demonstrate that the first eigenvector points strongly in the directions of $A_4$ and $A_6$, and slightly in the direction of $A_2$. Therefore, the uncertainty in the regressed values of these coefficients will be greatest in general, and their values correlated. From these radar plots, we also observe that the principal eigendirections are quite similar for the three experimental protocols considered. Certain features of the material response are therefore inherently more difficult to distinguish precisely in medium-amplitude, simple shear experiments, regardless of the type of data collected.

The uncertainty analysis presented in this section has revealed a few key features of the model identification problem posed in this work. Firstly, it is possible to perform model identification using a variety of data sets collected with different simple shear deformation protocols. However, not all of these data sets produce an equally well-conditioned regression problem. Therefore, even slight errors in experimental data from, for example, a startup of steady shear experiment, may result in large uncertainties in the regressed coefficients for the model. The condition number governing the propagation of relative uncertainty in the linear regression problem cannot be decreased arbitrarily by collecting more data; therefore, the relative uncertainty in the regressed coefficients must be controlled by thoughtful experimental design. In the end, however, each experimental protocol has inherent limits on its ability to decrease the condition number associated with the regression problem. Thus, experiments such as three-tone MAPS frequency sweeps, which produce a more diverse data set than single-tone MAOS frequency sweeps, are ultimately able to produce more well-conditioned regression problems, and often can do so with substantially less experimentation. Finally, though we have focused here on the simple linear regression problem relevant to the quadratic Maxwell model, we should expect the same observations to hold qualitatively for nonlinear problems, such as regularized regression or the nonlinear parameter estimation caused the parameter $\zeta$ in the cubic Maxwell model (equation \ref{eq:cubic_model}).

\section{Conclusions}

This work develops a general framework for Maxwell-Oldroyd type constitutive models, which we call the generalized nonlinear Maxwell model. Expansions of this model in weakly nonlinear flows, called the cubic and quadratic Maxwell models, each encompass the Oldroyd 8-constant framework, along with many other commonly applied constitutive models. The derived expressions for the third order complex viscosity in this framework, therefore, represents a library of solutions for the constitutive models contained within. We have demonstrated how this solution provides a convenient framework for model identification, which in the case of the quadratic Maxwell model can be reduced to a simple linear regression problem. Moreover, the mathematical framework of MAPS rheology makes it straightforward to extend the solution for the third order complex viscosity to medium amplitude expansions of the response of the model to specific simple shear deformations, such as single- and multi-tone oscillations and the startup of steady shear flow. This demonstrates the principal advantages of both MAPS rheology and such a general constitutive framework. The expression for the third order complex viscosity within this general framework arises naturally from an asymptotic expansion, and can be subsequently used to derive the third order complex viscosity and other medium amplitude response functions for all of the models contained within the generalized nonlinear Maxwell model. Thus, the need to carry out a full derivation for each model and each material function is avoided.

Another focus of this work has been to leverage the MAPS response of the generalized nonlinear Maxwell model to devise a simple quantitative framework for model identification. We have demonstrated that using the solution for the third order complex viscosity in this model as a fitting function for experimental data, one can simultaneously compare the optimal fit to a library of predictions by different constitutive models. Pairing this simple parameter estimation problem with a suitable regularization condition, such as the Bayesian information criterion employed by other authors \cite{ewoldt-2013}, results in a quantitative tool for assessing which of the many models captured within the generalized nonlinear Maxwell model best describes a particular complex fluid. In Section \ref{sec:model_identification}, we have demonstrated the model identification procedure explicitly using different types of medium amplitude rheological data, revealing that the choice of experimental protocol employed in this scheme can substantially influence one's ability to infer the parameters for the model with minimal uncertainty. In particular, the three-tone oscillatory shear experiments motivated by MAPS rheology outperform other well-known techniques such as MAOS and startup of steady shear, both in terms of the eventual relative uncertainty in the regressed parameters and in the amount of experimentation needed to realize a well-posed regression problem.

The principal goal of this work has been to demonstrate by example the power of MAPS rheology in theoretical and experimental explorations of constitutive models, including for model identification. However, it is worthwhile to consider the merits of the constitutive framework that we have developed -- that is, the generalized nonlinear Maxwell model -- on its own. As Oldroyd notes in his 1950 paper \cite{oldroyd-1950}, ``a theory of rheological phenomena can [...] be built up in two stages. First, it is necessary to specify the rheological properties [...] of the material by means of a set of rheological equations of state.'' Much in the spirit of Oldroyd's seminal work, the generalized nonlinear Maxwell model may fulfill the first stage for a wide variety of materials. Moreover, although the generalized nonlinear Maxwell model is not the only constitutive model capable of describing a wide range of viscoelastic behaviors, it is one of the most general constitutive frameworks that takes the form of a compact set of differential equations. ``The second stage'' Oldroyd asserts, ''is the prediction of the behavior of the material in bulk [...] by making use of the equations of state, the stress equations of motion, and the equation of continuity''. The structure of the generalized nonlinear Maxwell model is especially useful in this second stage, as its governing differential equations can be integrated directly into computational fluid dynamics tools, which already contain the infrastructure for solving multidimensional, coupled partial differential equations. Thus, even if a viscoelastic material might be well-described by a physically-inspired model of a different form, describing the material's response using the generalized nonlinear Maxwell model, for instance by parameterizing equation \ref{eq:cubic_model} for such a material, provides a pathway for using existing open source tools to simulate the material response in complex flow fields.

Along with being well-suited for computational fluid dynamics software, the generalized nonlinear Maxwell model is also primed for data-driven applications such as scientific machine learning. For instance, the nonlinear function $F(\boldsymbol{\sigma},\boldsymbol{\dot\gamma})$ may be replaced by a neural network and this model trained on various sets of rheological data, creating what other authors have called a `universal differential equation' (UDE) \cite{rackauckas-2020}. While detailed applications of rheological UDEs (or `rUDEs', for short) have yet to be explored, they do provide an exciting opportunity. In particular, such a data-driven tool would require only stress and strain time-series data for training; the length of these signals is unconstrained, as is the deformation protocol that was used to generate the data. In principle, then, any rheological data could be translated into a form fit for training this rUDE, representing a novel and invaluable pipeline assimilating vastly different rheological data sets, which can ultimately produce highly accurate and portable modeling tools. Of course, when data-driven methods such as rUDEs are employed for rheological modelling, obtaining optimal data sets for training these models is crucial for computational efficiency. In light of the conclusion from the present study that MAPS rheology provides particularly well-suited data for the linear parameter estimation problem, we anticipate that the MAPS framework in conjunction with Oldroyd's generalized formulation of differential constitutive models will be critical in building the data sets needed to train these next-generation computational rheological tools.

\section*{Acknowledgements}

K.R.L. was supported by the U.S. Department of Energy Computational Science Graduate Fellowship program under Grant No. DE-SC0020347.

\bibliographystyle{ieeetr}
\bibliography{biblio.bib}

\begin{thebibliography}{10}

\bibitem{rivlin-1971}
R.~S. Rivlin and K.~N. Sawyers, ``Nonlinear continuum mechanics of viscoelastic
  fluids,'' {\em Annual Review of Fluid Mechanics}, vol.~3, no.~1,
  pp.~117--146, 1971.

\bibitem{bird-1976}
R.~B. Bird, ``Useful non-{N}ewtonian models,'' {\em Annual Review of Fluid
  Mechanics}, vol.~8, no.~1, pp.~13--34, 1976.

\bibitem{bird-1987}
R.~B. Bird, R.~C. Armstrong, and O.~Hassager, {\em Dynamics of Polymeric
  Liquids, Volume 1: Fluid Mechanics}.
\newblock John Wiley \& Sons, Inc., 2~ed., 1987.

\bibitem{bernstein-1963}
B.~Bernstein, E.~A. Kearsley, and L.~J. Zapas, ``A study of stress relaxation
  with finite strain,'' {\em Trans. Soc. Rheol.}, vol.~7, no.~1, pp.~391--410,
  1963.

\bibitem{oldroyd-1958}
J.~G. Oldroyd, ``Non-{N}ewtonian effects in steady motion of some idealized
  elastico-viscous liquids,'' {\em Proc. R. Soc. Lond. A}, vol.~245, no.~1241,
  pp.~278--297, 1958.

\bibitem{oldroyd-1950}
J.~G. Oldroyd, ``On the formulation of rheological equations of state,'' {\em
  Proc. R. Soc. Lond. A}, vol.~200, no.~1063, pp.~523--541, 1950.

\bibitem{freund-2015}
J.~B. Freund and R.~H. Ewoldt, ``Quantitative rheological model selection: Good
  fits versus credible models using {B}ayesian inference,'' {\em Journal of
  Rheology}, vol.~59, no.~3, pp.~667--701, 2015.

\bibitem{tschoegl-1989}
N.~W.~Tschoegl, {\em The Phenomenological Theory of Linear Viscoelastic
  Behavior—An Introduction}.
\newblock Springer-Verlag, Berlin, 01 1989.

\bibitem{lodge-1964}
A.~S. Lodge, {\em Elastic liquids : an introductory vector treatment of
  finite-strain polymer rheology}.
\newblock Academic Press, 1964.

\bibitem{soussou-1970}
J.~E. Soussou, F.~Moavenzadeh, and M.~H. Gradowczyk, ``Application of {P}rony
  series to linear viscoelasticity,'' {\em Transactions of the Society of
  Rheology}, vol.~14, no.~4, pp.~573--584, 1970.

\bibitem{park-1999}
S.~W. Park and R.~A. Schapery, ``Methods of interconversion between linear
  viscoelastic material functions. {P}art {I} — a numerical method based on
  {P}rony series,'' {\em International Journal of Solids and Structures},
  vol.~36, no.~11, pp.~1653 -- 1675, 1999.

\bibitem{park-2001}
S.~W. Park and Y.~R. Kim, ``Fitting {P}rony-series viscoelastic models with
  power-law presmoothing,'' {\em Journal of Materials in Civil Engineering},
  vol.~13, no.~1, pp.~26--32, 2001.

\bibitem{yamamoto-1971}
M.~Yamamoto, ``Rate-dependent relaxation spectra and their determination,''
  {\em Trans. Soc. Rheol.}, vol.~15, no.~2, pp.~331--344, 1971.

\bibitem{owens-2002}
R.~G. Owens and T.~N. Phillips, {\em Computational rheology}.
\newblock World Scientific, 2002.

\bibitem{dewitt-1955}
T.~W. DeWitt, ``A rheological equation of state which predicts
  non‐{N}ewtonian viscosity, normal stresses, and dynamic moduli,'' {\em
  Journal of Applied Physics}, vol.~26, no.~7, pp.~889--894, 1955.

\bibitem{johnson-1977}
M.~W. Johnson and D.~J. Segalman, ``A model for viscoelastic fluid behavior
  which allows non-affine deformation,'' {\em Journal of Non-Newtonian Fluid
  Mechanics}, vol.~2, no.~3, pp.~255 -- 270, 1977.

\bibitem{johnson-1981}
M.~W. Johnson and D.~J. Segalman, ``Description of the non-affine motions of
  dilute polymer solutions by the porous molecule model,'' {\em Journal of
  Non-Newtonian Fluid Mechanics}, vol.~9, no.~1, pp.~33 -- 56, 1981.

\bibitem{giesekus-1982}
H.~Giesekus, ``A simple constitutive equation for polymer fluids based on the
  concept of deformation-dependent tensorial mobility,'' {\em Journal of
  Non-Newtonian Fluid Mechanics}, vol.~11, no.~1, pp.~69 -- 109, 1982.

\bibitem{tanner-1977}
N.~{Phan Thien} and R.~I. Tanner, ``A new constitutive equation derived from
  network theory,'' {\em Journal of Non-Newtonian Fluid Mechanics}, vol.~2,
  no.~4, pp.~353 -- 365, 1977.

\bibitem{leonov-1992}
A.~I. Leonov, ``Analysis of simple constitutive equations for viscoelastic
  liquids,'' {\em Journal of Non-Newtonian Fluid Mechanics}, vol.~42, no.~3,
  pp.~323 -- 350, 1992.

\bibitem{ewoldt-2013}
R.~H. Ewoldt and N.~A. Bharadwaj, ``Low-dimensional intrinsic material
  functions for nonlinear viscoelasticity,'' {\em Rheologica Acta}, vol.~52,
  pp.~201--219, Mar 2013.

\bibitem{lennon-2020-1}
K.~R. Lennon, G.~H. McKinley, and J.~W. Swan, ``Medium amplitude parallel
  superposition {(MAPS)} rheology. {P}art 1: Mathematical framework and
  theoretical examples,'' {\em Journal of Rheology}, vol.~64, no.~3,
  pp.~551--579, 2020.

\bibitem{lennon-2020-2}
K.~R. Lennon, M.~Geri, G.~H. McKinley, and J.~W. Swan, ``Medium amplitude
  parallel superposition {(MAPS)} rheology. {P}art 2: Experimental protocols
  and data analysis,'' {\em Journal of Rheology}, vol.~64, no.~5,
  pp.~1263--1293, 2020.

\bibitem{gurnon-2012}
A.~K. Gurnon and N.~J. Wagner, ``Large amplitude oscillatory shear {(LAOS)}
  measurements to obtain constitutive equation model parameters: {G}iesekus
  model of banding and nonbanding wormlike micelles,'' {\em Journal of
  Rheology}, vol.~56, no.~2, pp.~333--351, 2012.

\bibitem{graham-2003}
R.~S. Graham, A.~E. Likhtman, T.~C.~B. McLeish, and S.~T. Milner, ``Microscopic
  theory of linear, entangled polymer chains under rapid deformation including
  chain stretch and convective constraint release,'' {\em Journal of Rheology},
  vol.~47, no.~5, pp.~1171--1200, 2003.

\bibitem{likhtman-2003}
A.~E. Likhtman and R.~S. Graham, ``Simple constitutive equation for linear
  polymer melts derived from molecular theory: Rolie–{P}oly equation,'' {\em
  Journal of Non-Newtonian Fluid Mechanics}, vol.~114, no.~1, pp.~1 -- 12,
  2003.

\bibitem{chen-2018}
R.~T.~Q. Chen, Y.~Rubanova, J.~Bettencourt, and D.~Duvenaud, ``Neural ordinary
  differential equations.''
\newblock arXiv:1806.07366, June 2019.

\bibitem{rackauckas-2020}
C.~Rackauckas, Y.~Ma, J.~Martensen, C.~Warner, K.~Zubov, R.~Supekar,
  D.~Skinner, A.~Ramadhan, and A.~Edelman, ``Universal differential equations
  for scientific machine learning.''
\newblock arXiv:2001.04385, Jan. 2020.

\bibitem{spencer-1958}
A.~J.~M. Spencer and R.~S. Rivlin, ``The theory of matrix polynomials and its
  application to the mechanics of isotropic continua,'' {\em Archive for
  rational mechanics and analysis}, vol.~2, no.~1, pp.~309--336, 1958.

\bibitem{marrucci-2003}
G.~Marrucci and G.~Ianniruberto, ``Flow-induced orientation and stretching of
  entangled polymers,'' {\em Philosophical Transactions of the Royal Society of
  London. Series A: Mathematical, Physical and Engineering Sciences}, vol.~361,
  no.~1805, pp.~677--688, 2003.

\bibitem{larson-1984}
R.~G. Larson, ``A constitutive equation for polymer melts based on partially
  extending strand convection,'' {\em Journal of Rheology}, vol.~28, no.~5,
  pp.~545--571, 1984.

\bibitem{gordon-1972}
R.~J. Gordon and W.~R. Schowalter, ``Anisotropic fluid theory: A different
  approach to the dumbbell theory of dilute polymer solutions,'' {\em
  Transactions of the Society of Rheology}, vol.~16, no.~1, pp.~79--97, 1972.

\bibitem{saengow-2017}
C.~Saengow, A.~J. Giacomin, and C.~Kolitawong, ``Exact analytical solution for
  large-amplitude oscillatory shear flow from {O}ldroyd 8-constant framework:
  Shear stress,'' {\em Physics of Fluids}, vol.~29, no.~4, 043101, 2017.

\bibitem{williams-1962}
M.~C. Williams and R.~B. Bird, ``Three‐constant {O}ldroyd model for
  viscoelastic fluids,'' {\em The Physics of Fluids}, vol.~5, no.~9,
  pp.~1126--1128, 1962.

\bibitem{denn-1971}
J.~S. Ultman and M.~M. Denn, ``Slow viscoelastic flow past submerged objects,''
  {\em The Chemical Engineering Journal}, vol.~2, no.~2, pp.~81 -- 89, 1971.

\bibitem{nomenclature}
``Official symbols and nomenclature of {T}he {S}ociety of {R}heology,'' {\em
  Journal of Rheology}, vol.~57, no.~4, pp.~1047--1055, 2013.

\bibitem{pipkin-1964}
A.~C. Pipkin, ``Small finite deformations of viscoelastic solids,'' {\em
  Reviews of Modern Physics}, vol.~36, pp.~1034--1041, 1964.

\bibitem{martinetti-2019}
L.~Martinetti and R.~H. Ewoldt, ``Time-strain separability in medium-amplitude
  oscillatory shear,'' {\em Physics of Fluids}, vol.~31, no.~2, 021213, 2019.

\bibitem{vrentas-1991}
J.~S. Vrentas, D.~C. Venerus, and C.~M. Vrentas, ``{Finite amplitude
  oscillations of viscoelastic fluids},'' {\em Journal of Non-Newtonian Fluid
  Mechanics}, vol.~40, no.~1, pp.~1--24, 1991.

\bibitem{booij-1966}
H.~C. Booij, ``{Influence of superimposed steady shear flow on the dynamic
  properties of non-Newtonian fluids},'' {\em Rheologica Acta}, vol.~5, no.~3,
  pp.~222--227, 1966.

\bibitem{saengow-2017-2}
C.~Saengow and A.~J. Giacomin, ``{Normal stress differences from {O}ldroyd
  8-constant framework: Exact analytical solution for large-amplitude
  oscillatory shear flow},'' {\em Physics of Fluids}, vol.~29, p.~121601, aug
  2017.

\bibitem{rolon-2009}
V.~H. Rol{\'o}n-Garrido and M.~H. Wagner, ``The damping function in rheology,''
  {\em Rheologica Acta}, vol.~48, no.~3, pp.~245--284, 2009.

\bibitem{yong-2020}
H.~Y. Song, H.~J. Kong, S.~Y. Kim, and K.~Hyun, ``Evaluating predictability of
  various constitutive equations for {MAOS} behavior of entangled polymer
  solutions,'' {\em Journal of Rheology}, vol.~64, no.~3, pp.~673--707, 2020.

\bibitem{ramlawi-2020}
N.~Ramlawi, N.~A. Bharadwaj, and R.~H. Ewoldt, ``{The weakly nonlinear response
  and nonaffine interpretation of the Johnson–Segalman/Gordon–Schowalter
  model},'' {\em Journal of Rheology}, vol.~64, pp.~1409--1424, oct 2020.

\bibitem{geri-2018}
M.~Geri, B.~Keshavarz, T.~Divoux, C.~Clasen, D.~J. Curtis, and G.~H. McKinley,
  ``Time-resolved mechanical spectroscopy of soft materials via optimally
  windowed chirps,'' {\em Phys. Rev. X}, vol.~8, 041042, Dec 2018.

\bibitem{giacomin-1993}
A.~J. Giacomin and J.~M. Dealy, ``Large-amplitude oscillatory shear,'' in {\em
  Techniques in Rheological Measurement} (A.~A. Collyer, ed.), pp.~99--121,
  Dordrecht: Springer Netherlands, 1993.

\bibitem{mckinley-2008}
R.~H. Ewoldt, A.~E. Hosoi, and G.~H. McKinley, ``New measures for
  characterizing nonlinear viscoelasticity in large amplitude oscillatory
  shear,'' {\em Journal of Rheology}, vol.~52, no.~6, pp.~1427--1458, 2008.

\bibitem{poungthong-2019}
P.~Poungthong, A.~J. Giacomin, C.~Saengow, and C.~Kolitawong, ``Series
  expansion for shear stress in large-amplitude oscillatory shear flow from
  {O}ldroyd 8-constant framework,'' {\em The Canadian Journal of Chemical
  Engineering}, vol.~97, no.~S1, pp.~1655--1675, 2019.

\bibitem{singh-2018}
P.~K. Singh, J.~M. Soulages, and R.~H. Ewoldt, ``Frequency-sweep
  medium-amplitude oscillatory shear {(MAOS)},'' {\em Journal of Rheology},
  vol.~62, no.~1, pp.~277--293, 2018.

\bibitem{rivlin-1957}
A.~E. Green and R.~S. Rivlin, ``The mechanics of non-linear materials with
  memory,'' {\em Archive for Rational Mechanics and Analysis}, vol.~1,
  pp.~1--21, Jan 1957.

\bibitem{saengow-2019}
C.~Saengow, A.~J. Giacomin, N.~Grizzuti, and R.~Pasquino, ``Startup steady
  shear flow from the {O}ldroyd 8-constant framework,'' {\em Physics of
  Fluids}, vol.~31, no.~6, 063101, 2019.

\end{thebibliography}

\end{document}